\documentclass[11pt,a4paper]{article}
\usepackage[utf8]{inputenc}
\usepackage[T5]{fontenc}
\usepackage{lmodern}
\usepackage{microtype}
\usepackage[margin=2.5cm]{geometry}
\usepackage{setspace}
\usepackage{graphicx}
\usepackage{booktabs}
\usepackage{array}
\usepackage{indentfirst}
\usepackage{placeins}

\usepackage{pdflscape}
\usepackage{caption}
\usepackage[hyphens]{url}
\usepackage{xcolor}
\usepackage[colorlinks=true,urlcolor=blue!60!black,linkcolor=black,citecolor=black]{hyperref}
\title{Building and calibrating a 3Di subgrid flood model for Hanoi, without a surveyed terrain model, drainage data, or rain gauges}

\author{Minh Hoa Bui and Mai Anh Tran\\[0.5em] \normalsize TECOTEC TECHNOLOGIES JOINT STOCK COMPANY, Hanoi, Vietnam\\ \normalsize minhhoa0386@gmail.com, maianh@tecotec.com.vn}

\date{}

\begin{document}

\maketitle

\begin{abstract}

Urban hydrodynamic flood models typically rely on a surveyed terrain model, an explicit pipe drainage network, and rain gauge measurements, none of which were available for the inundation following Typhoon Matmo in Mỹ Đình and Tây Mỗ, Hanoi, on 7 October 2025. The present study addresses this issue by documenting the construction and calibration of a 32.8 km\textsuperscript{2} 3Di subgrid model under severe data constraints, demonstrating how unmonitored catchments can be simulated using openly accessible sources. Specifically, a 1~m digital elevation model was systematically repaired for river and lake bathymetry, unsurveyed water bodies, road underpasses, and an excavation pit to re-establish essential overland flow paths. In addition, assigned infiltration rates across land-cover classes were utilized to stand in for the unrecorded subsurface sewer system, while half-hourly satellite precipitation from the IMERG Late product provided rainfall forcing for a closed two-dimensional domain. Model calibration was conducted against eight crowdsourced depth observations. The simulation results demonstrated that six observation points retained surface water and were closely aligned with the reported values within 2.2~cm. However, two peripheral locations along Tây Mỗ Road and Road 70 remained completely dry in the model. Because all eight points were utilized during parameter adjustment and no independent validation dataset was withheld, this correspondence reflects a calibration fit rather than an unconstrained predictive test.

\end{abstract}


\section{Introduction}

\subsection{Urban flooding in Hanoi}

Hanoi lies in the Red River Delta, where the lowland plains of the city stand 5 to 20~m above sea level (Nguyen et al., 2025). Floods of the river are an inherent characteristic of this geographic setting, and historically they have been linked to the fertility of delta farmland (Gourou, 1936; Lê Bá Thảo, 2009). Over the past two decades, climate change has exponentially increased the frequency of heavy precipitation events, which substantially adds to the pressure of dense urban settlement on low-lying ground across Southeast Asian cities (Luo et al., 2018). Recent severe weather events in the past 20 years clearly illustrate this vulnerability. For example, Super Typhoon Yagi struck northern Vietnam on 7 September 2024, resulting in 21 casualties and 32 locations with flood depths exceeding 0.5~m in Hanoi, with Chương Mỹ district remaining submerged under up to 1.5~m of water for several days (Nguyen et al., 2025). In addition, the flood of 30 October to 1 November 2008 reported that up to 1~m of water inundated major city streets and paralyzed municipal transport, resulting in 92 casualties across northern Vietnam and southern China, as well as economic damages exceeding 177 million USD (Luo et al., 2018). Luo and colleagues' calibration of an inundation model against nine photographed stations achieved an R\textsuperscript{2} of 0.6, demonstrating that qualitative depth estimates derived from media imagery have precedent in constraining Hanoi flood simulations, while also confirming that the goodness of fit attainable from crowdsourced visual evidence remains inherently modest.

The storm event of October 2025 provides a comparable case of severe urban inundation. Typhoon Matmo (designated as Vietnam's storm number 11) moved inland across Guangxi, China, and on the afternoon of 7 October, the Hanoi Department of Construction identified 23 completely impassable intersections among more than 100 recorded flooded locations (Tiền Phong, 2025). Crucially, Mỹ Đình and Tây Mỗ, two densely populated urban centers in western Hanoi, were among the most severely affected districts.

A numerical flood model is only as informative as the empirical observations that constrain its boundary conditions, and for this event such monitoring data were virtually nonexistent. Therefore, the central problem this study addresses is whether an urban hydrodynamic model can be rigorously calibrated against informal crowdsourced evidence, provided that explicit measurement uncertainties are assigned to each point and the resulting calibration fit is interpreted without overstating its predictive validity.

\subsection{Three data gaps and this study}

The present study evaluates the extent to which a 3Di subgrid model can be constructed and calibrated for Typhoon Matmo across Mỹ Đình and Tây Mỗ when terrain topography, subsurface drainage, and meteorological forcing are all subject to severe uncertainty.

2D urban hydrodynamic simulations require an accurate digital terrain model, a representation of sewer drainage capacity, and high-resolution rainfall forcing. While standard modeling literature establishes workflows for each component, applying these methods in data-sparse environments introduces three compounding challenges. First of all, terrain models must accurately capture the physical landscape as overland flow encounters it (Ming et al., 2025). Second, urban drainage must be incorporated either through explicit 1D pipe networks or through parameterized surface approximations (Li et al., 2020; Li et al., 2023). Finally, precipitation forcing is ideally derived from calibrated rain gauge networks (Roversi et al., 2024).

The initial obstacle stems from the terrain dataset. The acquired digital elevation model removes vegetation, buildings, and bridges, yet it omits bathymetric bed profiles for the water bodies, contains an artificial excavation pit at the a construction site reaching over 6~m below the surrounding ground surface, and preserves the elevated road deck of the underpass rather than the depressed roadway below, distorting overland flow corridors before calibration begins. In addition, the lack of comprehensive underground sewer network data precluded explicit pipe modeling and prevented drainage-benchmark verification. Finally, the precipitation volume is accurately unknown.

To resolve the identified data deficiencies, the DEM was corrected through four conditioning procedures, subterranean pipe drainage was parameterized via land-cover-specific infiltration rates, and hydraulic roughness alongside underpass drainage rates were calibrated against eight crowdsourced depth records with explicit uncertainty intervals. This approach of substituting sewer capacity with constant infiltration follows previous precedent (Wang et al., 2018), where steady infiltration rates represented urban drainage performance more effectively than proportional rainfall discounts.

Because primary datasets for urban flood modeling are frequently closed or prohibitively expensive to acquire (such as airborne LiDAR surveys; Vernimmen et al., 2019), this study demonstrates an alternative framework utilizing open and surrogate data: the 2017 national DEM, satellite precipitation, satellite land-cover classifications, and news report depths. Overall, this study offers five specific contributions, each accompanied by explicit operational limitations. First, it establishes an end-to-end 3Di subgrid modeling workflow in a major metropolitan area lacking surveyed terrain, sewer networks, and rain gauges. Second, it details four DEM conditioning steps (restoring river bathymetry, interpolating unsurveyed lake beds, opening occluded underpasses, and filling an artificial excavation pit). Third, it parameterizes subsurface drainage through land-cover infiltration rates. Fourth, it provides a transparent calibration against eight news-derived depth intervals. Because all eight observations were utilized during parameter calibration and no independent validation dataset was reserved, this correspondence represents a calibration fit rather than an unconstrained validation. Lastly, it presents an analytical investigation into the failure mechanisms of the two dry observation points alongside the structural boundaries of the method.

\section{Theoretical background}

\subsection{Terrain preparation and subgrid elevation conditioning}

A hydrodynamic flood model can route surface water only along the topological paths that its underlying digital elevation surface explicitly contains, meaning that the choice and pre-processing of the elevation model establish an upper bound on what subsequent parameter calibration can recover. Ming et al. (2025) categorized elevation datasets into three principal configurations: a digital surface model (DSM) that retains above-ground features such as buildings and vegetation, a digital terrain model (DTM) representing bare ground, and a hybrid surface that stamps buildings and flow-controlling infrastructure back into the bare-earth model. While a DTM can overlook substantial flow-deflecting structures such as building blocks and elevated highways, a DSM frequently over-represents minor surface roughness when discretized at coarse resolutions. Because none of these three raw configurations is inherently free of artifacts that distort overland flow corridors, previous studies targeting street-scale hydrodynamic accuracy invariably incorporate a terrain conditioning step, implemented either through automated algorithmic filtering or through targeted manual editing.

The algorithmic filtering approach is exemplified by Abdullah et al. (2012), who developed a raw LiDAR filtering routine designed to process buildings, elevated roadways, bridges, and riverbanks. In a Kuala Lumpur test case, their two-dimensional hydrodynamic model demonstrated improved correspondence with observed inundation extents and depths relative to earlier baseline filtering routines. Conversely, the manual editing approach is illustrated by Ming et al. (2025), who manually inspected a hybrid elevation dataset to remove overhead bridge decks and buildings that artificially obstructed streets and river channels, as well as by Su et al. (2022), who eliminated vegetation and viaduct structures from a UAV elevation model by applying a geometric elevation threshold. In a related formulation, Amini et al. (2026) addressed topographic omissions by manually inserting 49 missing building footprints (representing 3.2\% of 1,537 structures) into a coastal flood domain. Notably, these studies share an underlying assumption: they treat surface obstructions as features protruding above the terrain that must be stripped away, without addressing the inverse problem where an unconditioned surface model inadvertently occludes subterranean or depressed passages that convey storm runoff in reality.

This inverse condition has received direct empirical attention in only a limited number of investigations. Specifically, Wang et al. (2018) identified that their raw digital elevation model failed to resolve a 1.5~m elevation drop between street pavement and an underpass in Wallington, London, necessitating manual elevation adjustments prior to hydrodynamic simulation. Similarly, Loftis et al. (2014) integrated airborne LiDAR elevations with bathymetric surveys in a subgrid simulation of New York City during Hurricane Sandy, demonstrating that occluded highway underpasses artificially obstructed overland drainage paths. Correcting these underpasses alongside building footprints increased spatial correspondence with mapped FEMA flood boundaries from 75\% to 80\%. However, their evaluation relied on mapped regional inundation extents rather than localized depth gauges, indicating the general direction of the improvement rather than its precise vertical magnitude. Submerged channel and bathymetric geometry introduces a related constraint. Neal et al. (2021) emphasized that river bathymetry cannot be directly captured via remote sensing at catchment scales and demonstrated that the choice of channel estimation technique exerts substantial influence. In a Mozambique case study, substituting uniform flow assumptions with gradually varied flow theory reduced modeled five-year flood extents by 40\% and floodplain storage by 79\%. Nevertheless, their methodology operated at regional scales and inferred bathymetric profiles from hydraulic flow theory, whereas the catchment examined here comprises complex urban lakes and a modified river channel with surveyed ground points. To the authors' knowledge, no previous investigation has simultaneously resolved occluded highway underpasses, an artificial road deck, an excavation pit, and unsurveyed lake basins within a single urban modeling domain. Because the current study applies all four conditioning steps (section 4.2), the terrain preparation presented here resolves a compound set of topographic defects that existing literature addresses only in isolation.

\subsection{Drainage approximations and infiltration proxies}

Where comprehensive subsurface infrastructure inventories exist, hydrodynamic models couple a one-dimensional pipe network to the two-dimensional overland surface to simulate subterranean conveyance explicitly. When such pipe data are absent, the sewer network must be approximated through a surrogate loss term whose formulation dictates what additional empirical information is required from the modeler. Li et al. (2020) classified simplified drainage representations into two primary families: discounted rainfall, which reduces incoming precipitation before runoff reaches the ground surface, and fixed infiltration, which models drainage losses as an auxiliary infiltration rate. In a comparative evaluation of five drainage approximations, Ming et al. (2025) noted that operational frameworks such as the UK Environment Agency standard combine a 30\% rainfall reduction with a uniform 12~mm/h drainage deduction (Environment Agency, 2019). Furthermore, Wang et al. (2018) calibrated a two-dimensional urban model of Wallington using crowdsourced media observations, demonstrating that assigning constant infiltration rates (50~mm/h for green spaces and 40~mm/h for sealed surfaces) reproduced flood recession trajectories far more realistically than a proportional rainfall discount. The two approaches diverge fundamentally in hydraulic behavior, since a rainfall discount ceases the moment precipitation ends, whereas an infiltration loss continues to drain ponded water from surface depressions throughout the recession phase.

Alternative approximation frameworks subtract fluid mass directly at assumed inlet locations, which requires substantial ancillary data. Specifically, Li et al. (2020) applied a weir equation capped by pipe conveyance capacity at stormwater inlets, a technique requiring verified inlet coordinates and physical design parameters. Similarly, Li et al. (2023) subtracted mass exclusively within street cells and calibrated these losses against a fully coupled 1D-2D benchmark simulation, which presupposes the availability of a reference hydraulic model for at least one historical event. Furthermore, Apel et al. (2024) implemented capacity-based rules within the RIM2D model to represent soil infiltration on pervious ground and roof drainage into sewers, which again assumes known drainage design thresholds. Because neither inlet locations nor a calibrated 1D-2D reference model exist for western Hanoi, these advanced approximation techniques cannot be implemented without fabricating the detailed engineering inputs they require.

The closest operational precedent to the current study is provided by Amini et al. (2026). Facing an absence of sewer GIS records in Hoboken, New Jersey, they calibrated a single effective infiltration rate across paved surfaces against 138 crowdsourced inundation reports; their optimal rate of 19~mm/h aligned closely with the municipal sewer design standard of approximately 20~mm/h, supporting the validity of interpreting calibrated infiltration as a bulk proxy for drainage capacity. However, as the authors acknowledged, this aggregate parameter operates as a bulk sink term that cannot decouple gravity conveyance, mechanical pumping, sewer surcharging, or localized backwater ponding. The present study adopts this bulk interpretation within a 2D-only domain, but assigns distinct infiltration rates across six land-cover classes and calibrates them against localized depth observations carrying explicit uncertainty bounds. Because this bulk term inherently absorbs all unmodeled subsurface drainage processes, the fitted rates cannot be interpreted as true soil infiltration parameters, an operational boundary that is carried directly into the parameter reporting in section 4.4.

\subsection{Rainfall estimation without ground gauges}

Published urban hydrodynamic models that achieve high street-level accuracy are typically driven by observed ground rainfall. For instance, Su et al. (2022) utilized an hourly rain gauge record, Amini et al. (2026) applied a single municipal rain gauge uniformly across their modeling domain, and Ming et al. (2025) drove simulations using radar precipitation fields at 1~km spatial and 5-minute temporal resolution. In contrast, because no rain gauge time series was accessible for the October 2025 storm in western Hanoi, precipitation forcing in the present study relies entirely on IMERG Late (version 07B), a satellite product providing half-hourly precipitation estimates on a 0.1$^{\circ}$ grid (Huffman et al., 2023). Consequently, precipitation uncertainty introduces a primary limitation for the simulation, because satellite rainfall products frequently exhibit substantial volume discrepancies relative to ground measurements during extreme convective storms.

This observed deficit is consistent with established literature indicating that satellite precipitation products systematically underestimate extreme convective intensities responsible for urban pluvial inundation. Roversi et al. (2024) evaluated seven precipitation products against more than 1,200 rain gauges across Vietnam during the 2020 typhoon season, finding that both IMERG (in its Early and Final runs) and ERA5-Land failed to resolve peak rainfall intensities. Similarly, Xu et al. (2024) reported that IMERG version 06B systematically underestimates heavy rainfall accumulations at the localized spatial scales required for urban flood modeling. In an evaluation of Super Typhoon Yagi over Hanoi in 2024, Nguyen et al. (2025) compared IMERG-Early, CMORPH-RT, and PDIR-Now against 25 ground stations; while IMERG-Early captured the storm's temporal peak better than competing satellite products, its detection rates in initial storm impact zones remained between only 8\% and 12\%. Although these three evaluations demonstrate a consistent negative bias for extreme events, they differ in geographic focus, storm dynamics, and product versions. None of them evaluated the IMERG Late run of version 07B utilized here; consequently, none provides an empirical correction factor that could be reliably applied to the present series. The exact sensitivity of modeled flood depths to this satellite rainfall deficit therefore remains an open question that is not formally simulated in this study.

\subsection{Crowdsourced observations from news and social media}

Crowdsourced imagery and eyewitness reports have been applied both to delineate flood boundaries and to constrain hydrodynamic models, though the precision of the extracted data determines how effectively it can constrain numerical simulations. Fohringer et al. (2015) extracted flood inundation extents and rough water depths from geolocated social media photographs during the 2013 Dresden flood. Similarly, Rosser et al. (2017) integrated geotagged social media imagery with remote sensing and topographic data to estimate flood probabilities in Oxford, achieving an area under the curve (AUC) of 0.93 on test data. However, both studies focused purely on mapping inundation extents rather than calibrating a physical hydrodynamic solver. Extending this workflow, Yu et al. (2016) evaluated a city-scale hydrodynamic model of Shanghai against crowdsourced flood reports, demonstrating broad agreement in flood timing and spatial distribution; they concluded that rainfall accuracy and micro-topography represent the two most sensitive controls on model performance, directly reinforcing the terrain and precipitation challenges examined in sections 2.1 and 2.3. Nevertheless, their crowdsourced dataset recorded only binary flood presence rather than quantitative water depths.

Deriving quantitative water depth from informal media sources presents much greater technical difficulty. Alizadeh Kharazi and Behzadan (2021) estimated street flood depths from submerged traffic signs in photographs using computer vision, reporting a mean absolute error of 12.63 inches (32~cm), an error margin that exceeds the total flood depth observed along many urban roadways. In Hanoi, Luo et al. (2018) calibrated an inundation model against nine photographed locations by inferring water depth from visible reference objects, achieving an R\textsuperscript{2} of 0.6 without quantifying observational uncertainty intervals. Furthermore, Amini et al. (2026) optimized surface infiltration against 138 binary reports and evaluated the result against eight photographic depths assigned an uncertainty budget of $\pm$5~cm, but did not compute formal continuous error metrics. Taken together, these studies demonstrate that crowdsourced depth estimates provide valuable empirical constraints in data-scarce settings, yet their inherent observation errors have rarely been carried directly into quantitative model evaluations. To the authors' knowledge, no published urban modeling study has reported formal continuous error metrics while explicitly carrying uncertainty intervals on every crowdsourced depth observation. Section 5 establishes this metric framework, defining observation ranges so that discrepancies between simulated and reported depths can be interpreted against observation noise.

\subsection{Subgrid hydrodynamic modeling formulations}

The 3Di modeling system utilized in this study solves the two-dimensional shallow water equations across a quadtree computational grid while preserving fine-scale topographic variability beneath the computational cells. Specifically, the subgrid method integrates high-resolution bathymetry and roughness rasters across each cell's volume and flow interfaces; this formulation produces a nonlinear relationship between water volume and water level, enabling an accurate representation of wetting and drying transitions across complex topography (Casulli, 2009; Casulli \& Stelling, 2011; Platzek et al., 2015). Because mass storage and frictional resistance account for subgrid terrain pixels, fluid can convey through linear depressions and channels that are substantially narrower than the computational grid resolution. Van Dam et al. (2012) noted that this formulation achieves computing speeds several orders of magnitude faster than conventional fine-grid solvers, while Stelling (2012) described the quadtree refinement scheme that concentrates high spatial resolution along roads, channels, and steep gradients.

However, the foundational literature also identifies the structural limits of the subgrid method. Platzek et al. (2015) demonstrated that the subgrid scheme remains accurate only when unresolved sub-cell features do not introduce substantial localized energy losses or abrupt flow blockages. An occluded highway underpass represents an artificial blockage of this type: if the elevation raster records a continuous road deck across an underpass trench, the subgrid formulation perceives an impermeable barrier where water physically conveys below the road deck. Consequently, subgrid discretization does not diminish the necessity of terrain conditioning; on the contrary, it increases the sensitivity of the model to topographic errors, because an artificial blockage that subgrid integration cannot resolve is propagated into every computational cell that contains it.

\section{Description of the case study and available data}

\subsection{Study area and geographical setting}

The study area encompasses 32.8 km\textsuperscript{2} situated around Mỹ Đình and Tây Mỗ within western Hanoi.

\begin{figure}[!htbp]
\centering
\includegraphics[width=0.9\linewidth]{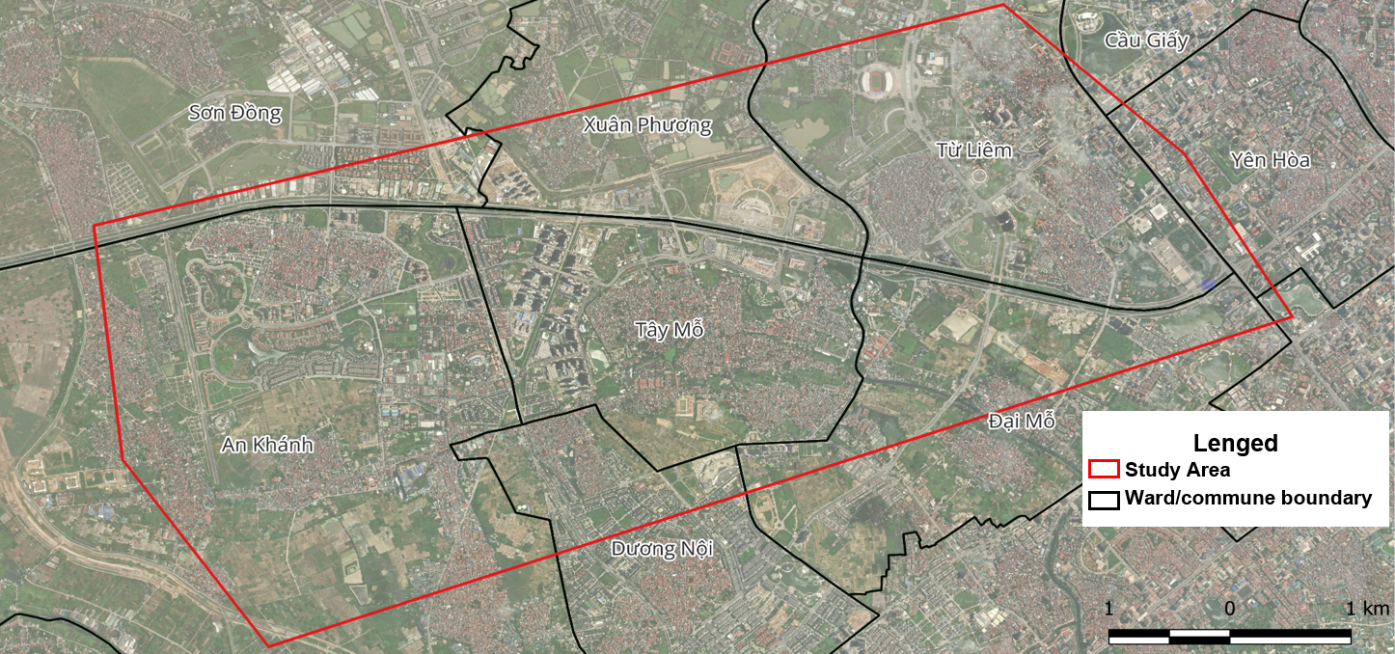}
\caption*{\textbf{Figure 1.} Study area (red outline) on a satellite basemap, with ward and commune boundaries (black lines).}
\end{figure}

\begin{figure}[!htbp]
\centering
\includegraphics[width=0.9\linewidth]{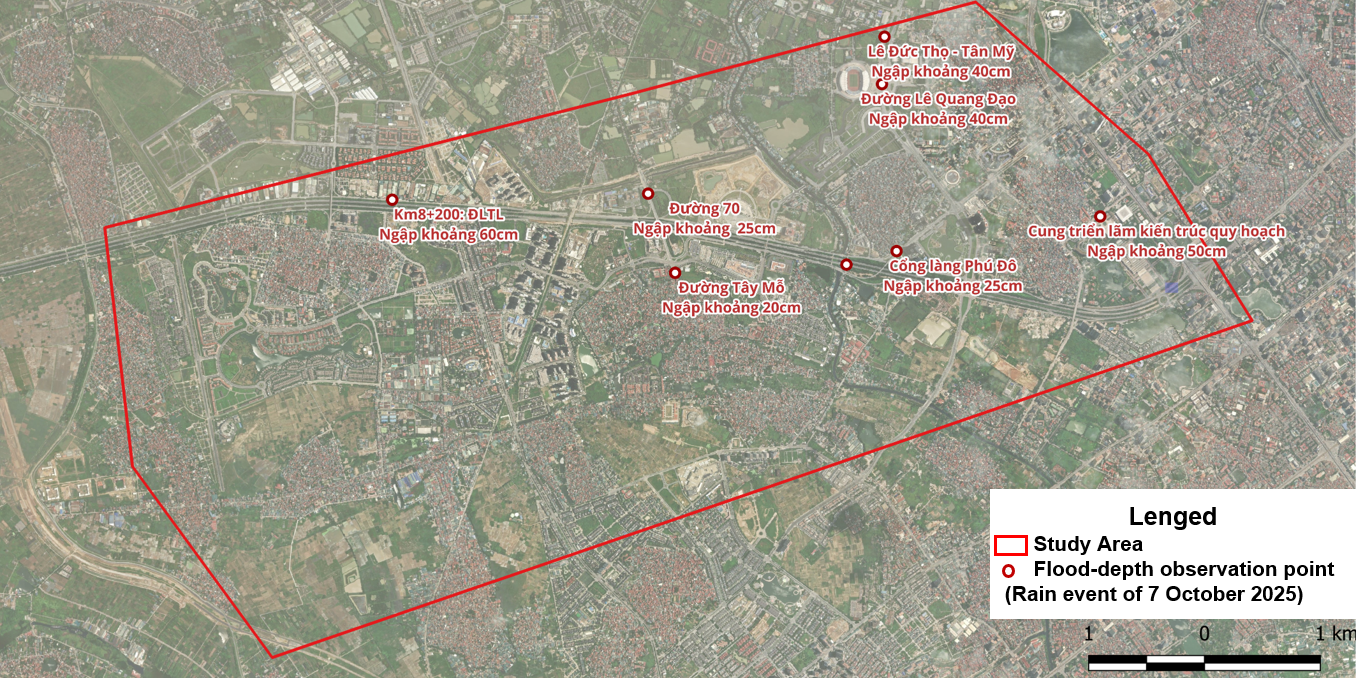}
\caption*{\textbf{Figure 2.} The depth observation points of 7 October 2025 with their reported depths.}
\end{figure}

The local hydrologic setting combines open surface water bodies with an extensive transportation infrastructure network that interrupts natural surface drainage pathways. Specifically, the Nhuệ River traverses the central modeling domain alongside several retention lakes such as Mễ Trì and Miếu Đầm, while Thang Long Boulevard carries major service corridors through sunken underpasses that serve as primary topographic collection sinks during extreme storms. Because two of the eight empirical depth observation points sit directly along the expressway corridor at Km4+944 and Km8+200, resolving underpass hydraulics is imperative for both local calibration and overall flood routing.

\subsection{Terrain and bathymetric survey data}

The underlying bare-earth surface is defined by a 1~m national digital elevation model surveyed in 2017 and provided by the Department of Survey, Mapping and Geographic Information of Vietnam, from which vegetation, buildings, and overhead bridges were removed. The resulting elevation grid contains 31.87 km\textsuperscript{2} of valid data cells referenced horizontally to VN-2000 / UTM zone 48N (EPSG:3405) with vertical elevations recorded in meters relative to Vietnam's national height datum at Hòn Dấu, Hải Phòng. In addition, all primary topographic inputs (including the base DEM and conditioned channel geometries), were projected to this vertical datum so that water levels could be evaluated consistently within the hydrodynamic solver.

Furthermore, because this terrain model captures topographic conditions from 2017 whereas the typhoon occurred eight years later, subsequent urban construction, land reclamation, and surface regrading are absent from the elevation matrix. Such unrecorded developments are most likely to misplace localized overland drainage channels and depression storage volumes within recently urbanized sub-catchments. Since project documentation provides no quantitative records of these topographical modifications, their hydraulic influence is carried forward as unquantified structural terrain error rather than being explicitly corrected.

A fundamental limitation of the raw digital elevation model is that optical surface surveys cannot penetrate open water surfaces to capture submerged bed topography. To address this discrepancy, an external survey contractor collected 17,386 bathymetric points utilizing single-beam sounding technology in March 2026, roughly five months following the flood event. However, this sounding campaign surveyed only the reach of the Nhuệ River, covering an area of roughly 4.2~km by 3.4~km, while omitting all interior retention lakes. Bathymetric depths for interior lakes and retention ponds had to be interpolated from peripheral shoreline elevations. The hydrodynamic model assumes that channel bed morphology remained stable between the flood event and the bathymetric survey.

\subsection{Land-cover classification}

Land cover classifications were derived from the ESA WorldCover 2021 dataset (version 200) at 10~m spatial resolution (Zanaga et al., 2022). Although eight distinct classes exist within the catchment boundary, shrubland and herbaceous wetland together occupy less than 0.001 km\textsuperscript{2}, leading us to retain six functional land categories in the hydrodynamic simulation. When resampled to the 1~m model grid, built-up surfaces account for 50.0\% of all computational cells, tree canopy covers 22.4\%, cropland occupies 14.9\%, grassland comprises 7.0\%, bare soil spans 3.8\%, and open water represents 1.8\%. Because impervious built-up terrain covers half of the total surface area, the infiltration rate assigned towards this class exerts primary control over domain-wide water retention. Notably, because the satellite classification reflects surface conditions from 2021, it omits land-use conversions occurring over the four intervening years, while its 10~m resolution cannot differentiate paved road corridors from permeable courtyards within individual urban parcels.

\subsection{Satellite rainfall forcing}

As local rain gauge records and sewer flow logs were formally requested but not provided, precipitation forcing was therefore constructed from the IMERG Late satellite product (version 07B), which generates half-hourly rainfall estimates across a 0.1$^{\circ}$ spatial grid of approximately 11~km resolution (Huffman et al., 2023).

Published ground observations provide a baseline to evaluate the magnitude of this satellite precipitation deficit. The national hydrometeorological forecasting center reported an accumulation of 255.5~mm at the Mễ Trì ground station between 19:00 on 6 October and 08:00 on 7 October 2025 (Chinhphu.vn, 2025), whereas the IMERG series recorded only 136.1~mm across the identical 13-hour window, representing a 47\% underestimation. Furthermore, municipal drainage agency bulletins published in local press cited cumulative totals up to 07:00 on 7 October of 207~mm at Nam Từ Liêm, 191~mm at Hà Đông, and 245~mm at Hoàng Cầu, without stating accumulation start times (Tuổi Trẻ, 2025). Across the corresponding interval from simulation start to 07:00, IMERG Late yielded 149.1~mm, falling 22\% to 39\% below these recorded drainage station figures. The same news source reported six-hour rainfall accumulations from the national forecasting center of 91~mm at Thượng Cát and 116~mm at Sóc Sơn without specifying end times, with the satellite estimate of 100.6~mm falling directly between these benchmarks.

None of these published ground benchmarks allows a direct cell-to-station verification. Each reported ground value reflects a localized point measurement from a gauge or pumping station, whereas the satellite figure represents an area-averaged mean across an 11~km grid cell where convective storm cells create pronounced spatial heterogeneity. Moreover, because these figures originate from secondary news summaries with unrecorded start times, discrepancies in accumulation windows may be due to reporting mismatches rather than pure satellite measurement error. Nevertheless, because the ground meteorological gauge and all municipal drainage station totals consistently exceed satellite accumulations, the satellite forcing must be regarded as an uncalibrated underestimate of true event precipitation.

\subsection{Storm event timeline and antecedent conditions}

Typhoon Matmo tracked inland across Guangxi, China, before weakening into a tropical depression by the morning of 6 October 2025, with meteorological agencies forecasting heavy convective rainfall over Hanoi starting from midday (VOV, 2025).

Prior hydrometeorological conditions from 25 September to 1 October 2025, which included the passage of Typhoon Bualoi, contributed significantly to the disaster by depositing 150 to 250~mm of widespread precipitation and localized totals exceeding 450~mm across Hà Nội, Phú Thọ, and Lào Cai (VietNamNet, 2025). Consequently, catchment soils and natural retention water bodies were already wet when Typhoon Matmo commenced, leading to elevated runoff coefficients and diminished storage capacity in urban depressions. The current hydrodynamic schematization captures only a fraction of these antecedent conditions. Specifically, elevated baseline stages within Nhuệ River and urban lakes are incorporated via a uniform initial water level raster applied at the start of simulation. However, as the model lacks a coupled unsaturated groundwater module, soil profiles are not modeled as pre-saturated, meaning that surface infiltration operates directly from the nominal loss rates assigned to each land-cover class.

\section{Methods}

\subsection{Hydrodynamic model and domain setup}

The hydrodynamic flood simulation utilizes the 3Di subgrid solver (van Dam et al., 2012, section 2.5), which solves the two-dimensional shallow water equations across a flexible quadtree grid while integrating fine sub-cell topography from the 1~m bare-earth DEM. The computational mesh encompasses 20,838 active cells over the 32.8 km\textsuperscript{2} domain, with cell dimensions ranging from 4~m to 64~m, with the finest 4~m and 8~m cells positioned within targeted refinement zones around highway underpasses, the Nhuệ River channel, and localized areas of steep topographic relief.

\begin{figure}[!htbp]
\centering
\includegraphics[width=0.9\linewidth]{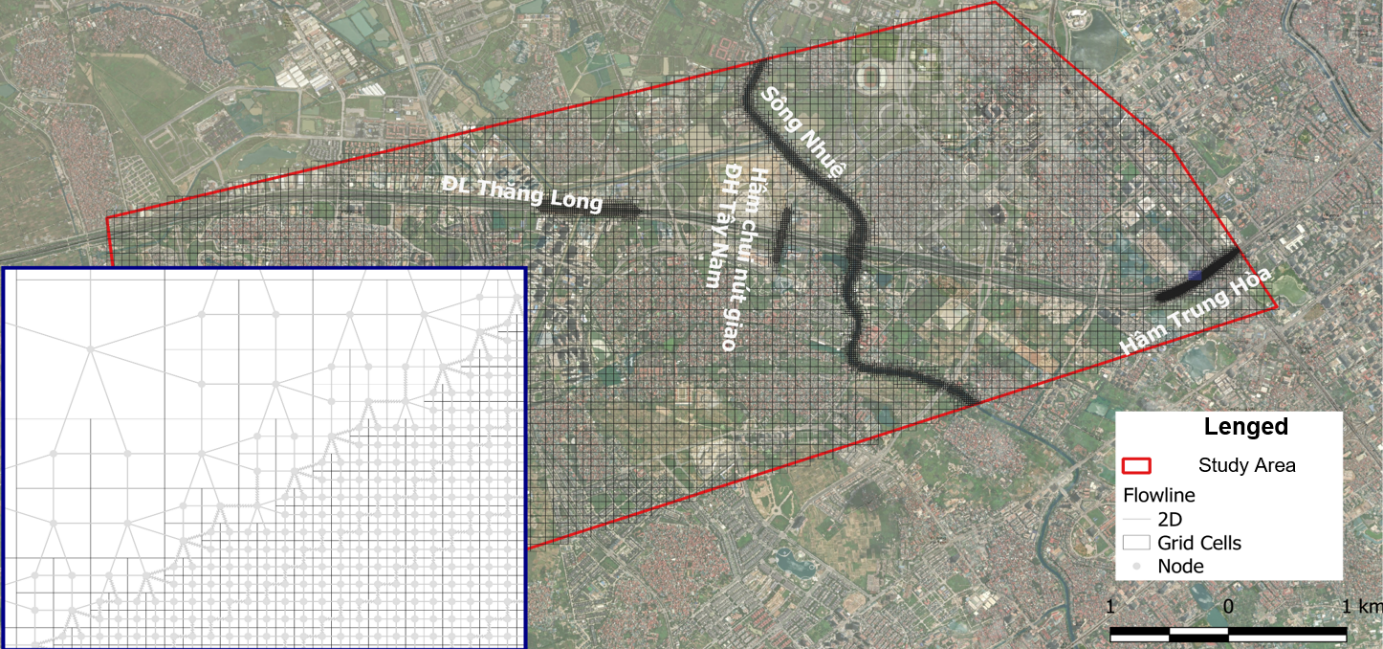}
\caption*{\textbf{Figure 3.} Computational grid of an earlier set-up of the model: base cell of 64~m, refined around the rivers and underpasses to 8~m (subgrid calculation on the 1~m DEM).}
\end{figure}

Notably, the computational domain operates as a hydrologically closed system without boundary conditions. The schematization incorporates neither subsurface drainage conduits, mechanical pump stations, external boundary stages, nor lateral inflows; thus, direct rainfall provides the sole water inflow while surface infiltration constitutes the only outflow mechanism. While this closed architecture isolates inundation produced by rainfall falling strictly within the catchment boundary, it simultaneously excludes external riverine inflow from the upper Nhuệ River and conveyance through the municipal sewer network, meaning that flood depths influenced by either factor cannot be reproduced by the simulation. Rainfall is distributed uniformly across all computational cells as two-dimensional surface rain. Bed friction is parameterized through the Manning formulation and infiltration losses are calculated using the 3Di simple infiltration module, with both parameters supplied as spatial rasters. The hydrodynamic simulation runs continuously for 65 hours.

\subsection{Digital elevation model conditioning}

The raw 2017 national elevation model contains four systematic structural defects that artificially obstruct or misdirect overland runoff, each of which was resolved through targeted raster pre-processing prior to model execution (see Appendix B). In total, these modifications affected 77 distinct contiguous zones comprising at least 200 cells each, altering 1.84 km\textsuperscript{2} or 5.8\% of the valid elevation data cells, where an individual cell was defined as edited when its vertical elevation changed by more than 0.02~m. However, these conditioning steps vary substantially in their supporting empirical evidence. Only the river bathymetry repair relies on direct in-situ sounding measurements, whereas the remaining three procedures depend on geometric interpolation or qualitative engineering judgment, which limits how far the modified surface can be considered an exact representation of actual ground topography.

\begin{figure}[!htbp]
\centering
\includegraphics[width=0.9\linewidth]{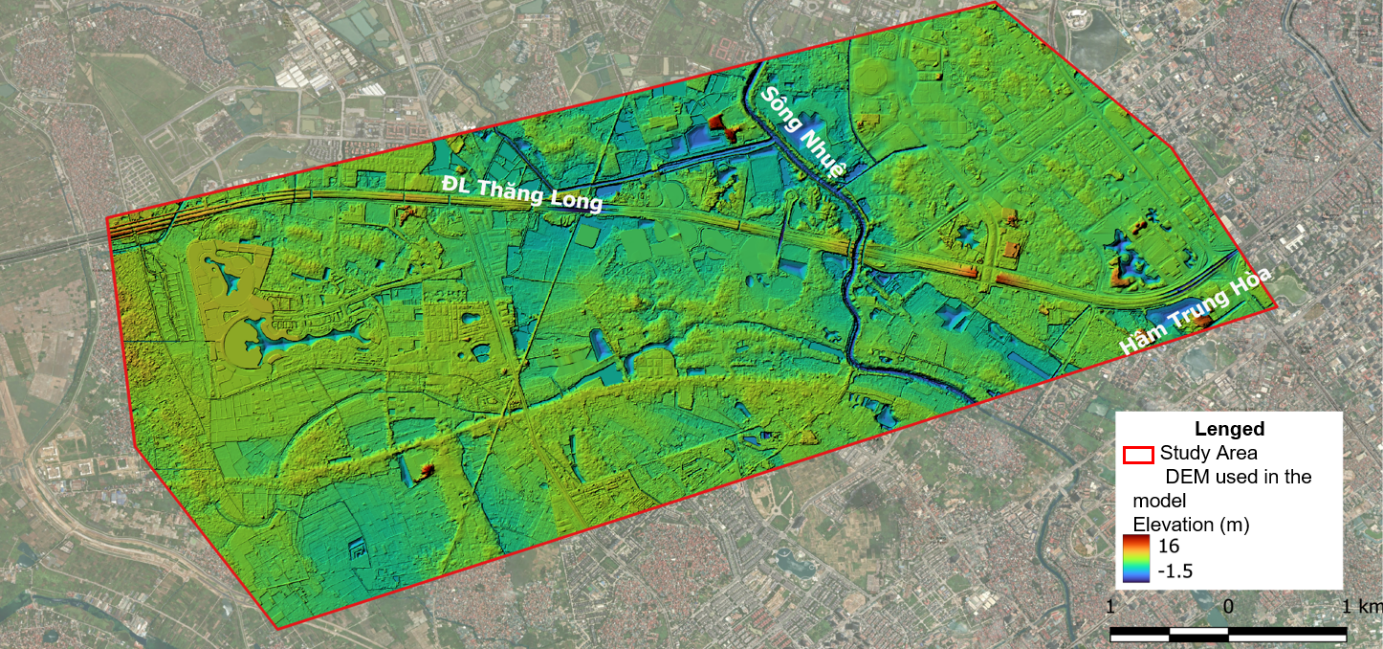}
\caption*{\textbf{Figure 4.} The DEM used in the model, with the Nhuệ River, Thang Long Boulevard, and the Trung Hòa underpass.}
\end{figure}

\subsection{Meteorological input processing}

A single spatially lumped precipitation hyetograph drives the hydrodynamic simulation, as the computational domain remains small relative to the resolution of the satellite product. This precipitation series was constructed by computing an area-weighted average across the four IMERG Late grid cells that intersect the spatial bounding box of the model domain, converted into incremental rainfall depths for each 30-minute time step. Since the modeling domain is substantially smaller than a single satellite cell footprint, the resulting hyetograph contains no spatial variability, meaning that real micro-scale precipitation gradients between Mỹ Đình and Tây Mỗ cannot be captured. The 131 half-hourly intervals yield a cumulative precipitation depth of 250.69~mm, with a maximum 30-minute peak increment of 28.24~mm (equivalent to an intensity of 56.5~mm/h), commencing synchronously with the simulation on 5 October 2025 at 17:00 UTC. Precipitation forcing represents the sole source of mass inflow into the closed computational domain. As a result, systematic measurement errors in the satellite hyetograph propagate directly into modeled surface water depths.

\begin{figure}[!htbp]
\centering
\includegraphics[width=0.9\linewidth]{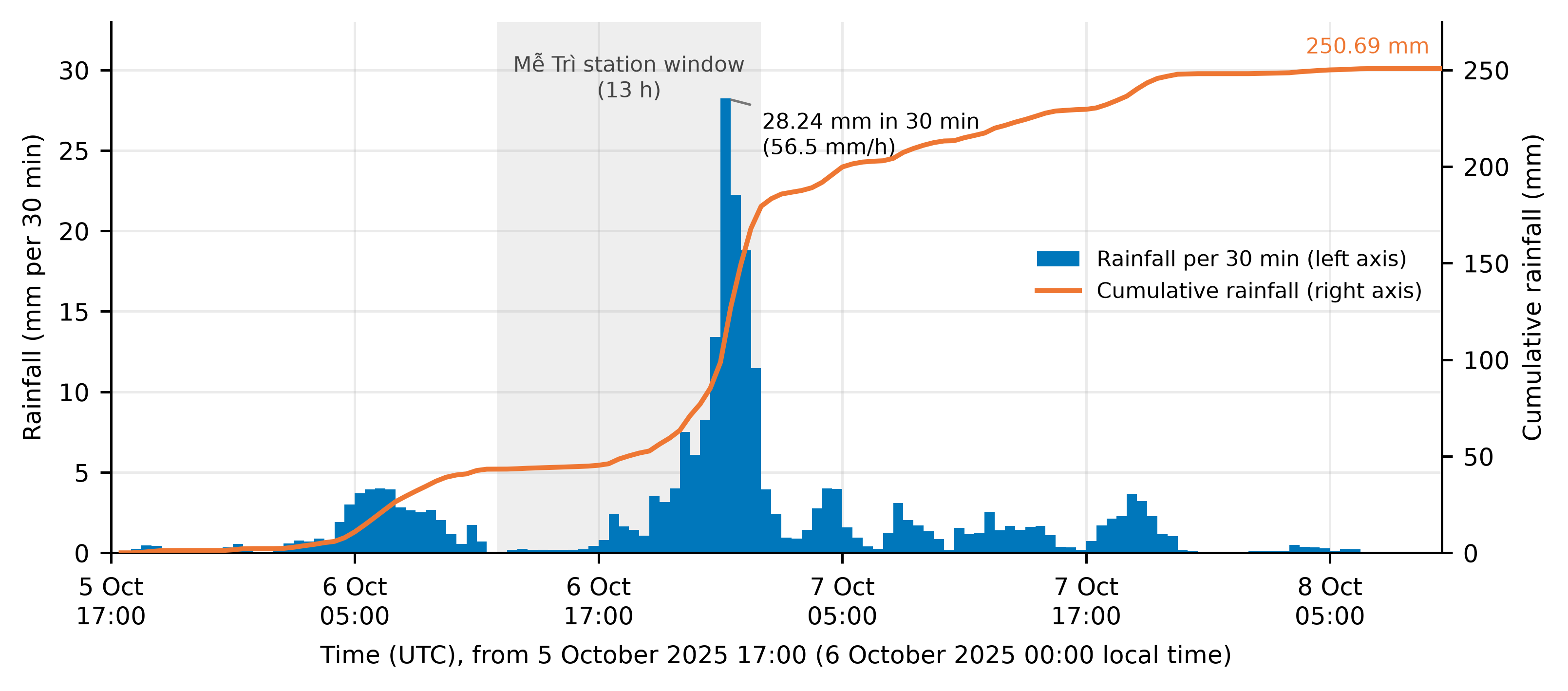}
\caption*{\textbf{Figure 5.} Rainfall series that drives the model, in mm per 30 minutes (bars) with the cumulative total (orange line).}
\end{figure}

\subsection{Surface roughness and infiltration parameterization}

Hydraulic roughness parameters were specified according to surface features, while infiltration loss rates were assigned by land-cover classification, with calibrated parameters. Initial baseline values for Manning's n were established based on standard literature references. Specifically, starting values for paved roadways (0.013), bare ground (0.035), and the natural channel of Nhuệ River (0.035) fall directly within the empirical roughness intervals compiled by Chow (1959) for smooth asphalt, fallow cultivated soil, and clean natural streams. However, Chow (1959) provides no direct roughness classifications for urban residential blocks or open lakes, and the selected baseline roughness of 0.060 for agricultural parcels exceeds standard values for row crops, meaning that these parameters relied primarily on professional estimation. Initial infiltration parameters were established from empirical Horton ranges across each ESA WorldCover land category. Subsequent calibration adjustments modified ten of the twelve adjustable hydraulic parameters. The background roadway roughness (0.013) and the agricultural roughness (0.060) remained at their initial settings, whereas the infiltration capacity for tree cover was calibrated to 20~mm/h, matching the upper boundary of its initial range.

In addition, ten localized spatial patches situated along Thang Long Boulevard were assigned elevated infiltration capacities of 300 or 500~mm/h. Infiltration rates of this magnitude rapidly evacuate surface water directly from the expressway pavement rather than routing it into vulnerable sunken underpasses, thereby acting as a localized surrogate for the missing stormwater drainage infrastructure.

Parameterized infiltration simultaneously serves as an operational proxy for the unmapped subsurface sewer network, which carries fundamental implications for hydraulic interpretation. Specifically, the calibrated rate of 1~mm/h assigned across built-up land (which represents 50.0\% of the modeling domain) reflects bulk drainage abstraction into subterranean pipes rather than genuine soil infiltration, which impervious asphalt and concrete surfaces largely prohibit. This value is substantially lower than effective drainage rates of 6 to 40~mm/h reported in comparable urban hydrodynamic modeling studies (Ming et al., 2025; Amini et al., 2026; Wang et al., 2018). The difference is because those previous investigations incorporated dedicated subterranean sewer networks or calibrated higher bulk capacity allowances, the minimal 1~mm/h loss rate utilized here ensures that nearly all precipitation falling on impervious surfaces is retained on the surface. The initial baseline range for built-up surfaces was set at 0 to 0.5~mm/h, with calibration marginally increasing it to 1~mm/h. Due to the likelihood that the satellite precipitation series systematically underestimates storm volume, this depressed drainage loss rate on urban surfaces may compensate for the rainfall deficit by artificially preserving ponded water volumes.

\begin{table}[!htbp]
\centering
\setlength{\tabcolsep}{6pt}
\caption*{Table 1. Final parameter values.}
\begin{tabular}{>{\raggedright\arraybackslash}p{0.202\dimexpr\textwidth-60pt\relax}>{\raggedright\arraybackslash}p{0.383\dimexpr\textwidth-60pt\relax}>{\raggedright\arraybackslash}p{0.117\dimexpr\textwidth-60pt\relax}>{\raggedright\arraybackslash}p{0.149\dimexpr\textwidth-60pt\relax}>{\raggedright\arraybackslash}p{0.149\dimexpr\textwidth-60pt\relax}}
\toprule
Parameter & Class or feature & Final value & Before calibration & Share of cells \\
\midrule
Manning's n & Background, including roads & 0.013 & 0.013 & 54.28\% \\
Manning's n & Bare land, construction sites, and others & 0.036 & 0.035 & 29.16\% \\
Manning's n & Urban and residential areas, farm land & 0.060 & 0.060 & 13.42\% \\
Manning's n & Lakes and ponds & 0.020 & 0.025 & 2.67\% \\
Manning's n & Nhuệ River and channels & 0.025 & 0.035 & 0.47\% \\
Infiltration (mm/h) & Tree cover & 20 & 10 to 20 & 22.4\% \\
Infiltration (mm/h) & Grassland & 15 & 5 to 10 & 7.0\% \\
Infiltration (mm/h) & Cropland & 10 & 2 to 5 & 14.9\% \\
Infiltration (mm/h) & Built-up & 1 & 0 to 0.5 & 50.0\% \\
Infiltration (mm/h) & Bare vegetation & 60 & 1 to 3 & 3.8\% \\
Infiltration (mm/h) & Water & 0 & 0 & 1.8\% \\
Infiltration (mm/h) & Underpass and road trench patches & 300 and 500 & more than 100 & 0.16\% \\
Infiltration (mm/h) & Default where the raster has no value & 50 & 50 & not applicable \\
\bottomrule
\end{tabular}
\end{table}

\subsection{Initial boundary stage specification}

A two-dimensional initial water level raster imposes a uniform stage of 3.5~m across 165,101 grid cells, encompassing 0.165 km\textsuperscript{2} of the Nhuệ River corridor and interior lake basins. This elevation was extracted from Google Earth spot elevations to reflect elevated antecedent surface stages produced by the heavy rainfall spells of late September.

\subsection{Depth observation collection and calibration metrics}

\subsubsection{Collecting opportunistic depths from news media}

Crowdsourced estimates published by media agencies constitute the sole empirical data source for model calibration. Eight localized flood depth observations form the final calibration dataset (see Appendix A). Specifically, six points were retrieved from a detailed Tuổi Trẻ investigative report and two points were extracted from a municipal bulletin listing 23 flooded locations compiled by the Hanoi Department of Construction and published in Tiền Phong. Since six of the eight observation sites originate from a single journalist, they inevitably reflect that correspondent's specific spatial itinerary and reporting timeline, meaning that they cannot be treated as eight independent samples of the flood. To construct this dataset, the authors initially screened 68 localized flood reports compiled across five digital news outlets, retaining only those points that fell inside the computational domain and provided either photographic documentation or explicit textual depth figures. Four of the eight selected locations include photographic evidence, while the remaining four points rely on quantitative figures stated directly in published news text that were verified against original reporting.

Geographic coordinates for each observation point were derived directly from topographic place names cited in the news reports. Because news reports do not publish spatial coordinates, the authors identified each named site in Google Earth by matching local landmark features between report imagery and satellite views, subsequently verifying the candidate position using three-dimensional perspective tools. These spatial coordinates therefore represent manual inferences by the authors, and their exact horizontal positioning error was not formally measured. The observation timestamp for each location corresponds to the clock time explicitly cited in the article text or photo caption. Where the source omitted a specific timestamp, the time of article publication was adopted as a proxy, and where reports provided only a broad diurnal window such as ``afternoon'', a standardized time of 15:00 was assigned. Notably, because raw photograph metadata was inaccessible across all digital articles, stated or published publication times may diverge from the actual instant of photography, introducing unquantified temporal observation uncertainty into the calibration sequence.

Inundation depths were derived using two distinct methodological approaches. Four depths (P01, P02, P03, and P06) were extracted visually from photographs, whereas the remaining four figures were stated directly in published text, comprising two journalist estimates (P04 and P05) and two official agency figures (P07 and P08). For photographic observations, the authors identified a visible reference object submerged in the water, such as electric motorbike wheels, passenger car tires, commuter bicycles, or pedestrians wading through the flood, estimating the submerged vertical fraction and multiplying this proportion by typical physical dimensions of the object. None of the physical reference objects was directly measured in the field. Object dimensions were assumed based on standard specifications for the most widely operated electric motorbikes and compact passenger cars in Hanoi, alongside mean anthropometric height data for adult Vietnamese populations to determine shin, knee, and thigh water levels. While text figures were incorporated exactly as reported, photographic interpretations fundamentally depend on assumed object geometry and subjective waterline identification, whereas stated journalist figures provide no verifiable method, which may be due to a lack of reporting standards. Furthermore, visual scale references are weak in certain instances: the photograph for P01 contains no clear calibration object in the water, while the image for P06 captures the receding edge of the flooded corridor rather than its maximum center depth.

Each observation entry incorporates a minimum, a central, and a maximum depth bound in centimeters, in order to evaluate hydrodynamic simulation accuracy against an explicit tolerance window rather than an idealized scalar value. Visual photographic readings were assigned a nominal uncertainty interval of $\pm$10~cm, whereas explicitly stated text figures were budgeted at $\pm$5~cm, with original reported ranges preserved wherever provided by the source. These uncertainty bounds exist because primary journalistic sources provide no technical description of how depths were measured or estimated. The narrower tolerance for stated figures reflects the assumption that reporters and municipal officials round estimates to convenient multiples, whereas the wider bound for photographic readings accounts for variance in object sizes and uncertain waterline positioning. Amini et al. (2026) similarly budgeted $\pm$5~cm for crowdsourced flood depths derived from media photographs, which supports the general magnitude of this error budget, if not the precise numerical thresholds adopted here.

One specific observation point departs from this standardization scheme, which directly influences aggregate error metrics. For P02, two independent visual readings of the photograph yielded conflicting depths of 25~cm and 50~cm, producing a composite observation range of 25 to 50~cm. The central value was designated as 50~cm, based on pedestrians wading across the building forecourt with water reaching between the shin and knee level, meaning that this central benchmark sits at the extreme upper boundary of its own uncertainty interval. Similarly, the reported range for P06 (15 to 40~cm) was retained directly from article text, although the standard photographic rule would have yielded an interval of 15 to 35~cm.

Every candidate entry was additionally assigned a qualitative confidence grade of high, medium, or low as a means of ranking data reliability. Specifically, a high confidence rating requires an unambiguous submerged scale object or a documented measurement method, a verified observation timestamp, and no internal discrepancies between independent readings. A medium rating indicates a single photographic reading with an imperfect scale reference, or a stated textual figure lacking an explicit timestamp or measurement methodology. Finally, a low rating denotes conflicting interpretations or photographs lacking clear vertical reference objects.

\subsubsection{Calibration points and statistical error metrics}

The eight observation points P01 through P08 were utilized directly to tune model parameters. Calibration adjustments modified up to twelve parameters: five Manning's n roughness coefficients, five land-cover infiltration capacities, and two localized underpass infiltration rates. With twelve adjustable numerical parameters tuned against only eight empirical targets, an acceptable mathematical fit is naturally expected from optimization alone, so agreement across these points cannot distinguish between a robust physical model and an over-calibrated parameterization. The present study reports an empirical calibration fit without claiming independent predictive capability.

The wide uncertainty intervals assigned towards observed depths impose a second fundamental constraint on physical inference. The observational bounds of $\pm$5~cm or $\pm$10~cm substantially exceed a residual difference of 2~cm; therefore, a modeled depth falling within 2~cm of the central observed benchmark cannot be differentiated from observational measurement noise. Consequently, minimal error residuals demonstrate mathematical consistency with informal crowdsourced reports, but they do not prove that the simulation has captured true physical inundation depths.

Performance evaluation is conducted using surface water depth rather than absolute water stage. As the modeled water depths and underlying ground elevations are derived from the identical digital elevation raster, depth comparisons are completely independent of vertical datum transformations. Simulated depths at each observation timestamp were calculated via linear temporal interpolation between the two nearest 30-minute output intervals of the 3Di water depth raster. Any observation point that the hydrodynamic model leaves completely dry is assigned a simulated water depth of 0~cm.

\section{Results}

\subsection{Model calibration performance}

Table 2 compares simulated water depths at each of the eight calibration locations against their corresponding central reported values recorded at the specific observation timestamps. As these eight empirical targets were directly utilized to tune the numerical parameters, this evaluation quantifies how closely the calibrated model reproduces its training targets, meaning that it provides no evidence regarding predictive performance at unmonitored locations.

\begin{table}[!htbp]
\centering
\small
\setlength{\tabcolsep}{3pt}
\caption*{Table 2. Observed and simulated depths in the final run.}
\begin{tabular}{>{\raggedright\arraybackslash}p{0.054\dimexpr\textwidth-48pt\relax}>{\raggedright\arraybackslash}p{0.321\dimexpr\textwidth-48pt\relax}>{\raggedright\arraybackslash}p{0.098\dimexpr\textwidth-48pt\relax}>{\centering\arraybackslash}p{0.089\dimexpr\textwidth-48pt\relax}>{\raggedright\arraybackslash}p{0.107\dimexpr\textwidth-48pt\relax}>{\centering\arraybackslash}p{0.116\dimexpr\textwidth-48pt\relax}>{\centering\arraybackslash}p{0.116\dimexpr\textwidth-48pt\relax}>{\raggedright\arraybackslash}p{0.098\dimexpr\textwidth-48pt\relax}}
\toprule
ID & Place & Observed range (cm) & Central (cm) & Simulated (cm) & Difference (cm) & Difference (\%) & Time compared \\
\midrule
P01 & Tây Mỗ Road & 10 to 30 & 20 & 0 (dry) & $-$20 & $-$100 & 08:20 \\
P02 & National Exhibition Center for Planning and Architecture & 25 to 50 & 50 & 50.1 & +0.15 & +0.3 & 09:20 \\
P03 & Phú Đô village gate & 15 to 35 & 25 & 23.2 & $-$1.82 & $-$7.3 & 08:20 \\
P04 & Lê Đức Thọ- Tân Mỹ Street & 35 to 45 & 40 & 38.9 & $-$1.08 & $-$2.7 & 08:17 \\
P05 & Lê Quang Đạo Street & 35 to 45 & 40 & 38.8 & $-$1.19 & $-$3.0 & 08:17 \\
P06 & Road 70 & 15 to 40 & 25 & 0 (dry) & $-$25 & $-$100 & 09:20 \\
P07 & Km8+200, Thang Long Boulevard & 55 to 65 & 60 & 62.2 & +2.17 & +3.6 & 15:00 \\
P08 & Km4+944 underpass & 95 to 105 & 100 & 98.4 & $-$1.63 & $-$1.6 & 15:00 \\
\bottomrule
\end{tabular}
\end{table}

The simulation results demonstrated that six of the eight observation sites held surface water, with simulated depths matching central reported values within 2.2~cm at every wet location. However, at P01 and P06 the model predicted zero inundation across all 15 output intervals, meaning that the depth raster contains no water values at these coordinates; hence both points were evaluated with a simulated depth of 0~cm.

\begin{table}[!htbp]
\centering
\setlength{\tabcolsep}{6pt}
\caption*{Table 3. Metrics of the final run.}
\begin{tabular}{>{\raggedright\arraybackslash}p{0.397\dimexpr\textwidth-36pt\relax}>{\raggedright\arraybackslash}p{0.167\dimexpr\textwidth-36pt\relax}>{\raggedright\arraybackslash}p{0.436\dimexpr\textwidth-36pt\relax}}
\toprule
Metric & All eight points & Six points with water \\
\midrule
Bias (cm) & $-$6.05 & $-$0.57 \\
Mean absolute error (cm) & 6.63 & 1.34 \\
RMSE (cm) & 11.39 & 1.49 \\
Mean absolute difference (\%) & 27.3 & 3.1 \\
Largest difference & $-$25~cm at P06 & +2.17~cm at P07 and $-$7.3\% at P03 \\
Share within 10~cm & 6 of 8 & 6 of 6 \\
Share inside the observed range & 5 of 8 & 5 of 6 (P02 is 0.1~cm above 50~cm) \\
\bottomrule
\end{tabular}
\end{table}

Crucially, aggregate statistical error metrics are governed primarily by the two dry failure points rather than the six successfully calibrated locations. Excluding P01 and P06 from the sample lowers the mean absolute error from 6.63~cm to 1.34~cm, reduces the root mean square error from 11.39~cm to 1.49~cm, and attenuates mean bias from $-$6.05~cm to $-$0.57~cm. The substantial disparity between RMSE and MAE across all eight points (11.39~cm versus 6.63~cm) confirms that residual variance is concentrated in isolated severe errors rather than distributed evenly across the modeling domain. Furthermore, the persistent negative bias observed in both groupings indicates that the hydrodynamic model systematically underpredicts surface water accumulation relative to eyewitness accounts. Because reported observational uncertainty intervals span 10 to 25~cm, the 1.34~cm mean absolute error across wet locations sits well within the measurement bounds of informal crowdsourced records.

Observational timing introduces a critical sensitivity across morning observation locations because simulated water levels exhibited rapid recession during the event window. Specifically, between 08:00 and 14:00, simulated water depths receded from 25.5 to 11.2~cm at P03, from 40.2 to 30.2~cm at P04, and from 39.9 to 29.2~cm at P05. Consequently, a discrepancy of merely 30 minutes in the reported observation timestamp alters simulated water depth by 2 to 4~cm, which ironically, may contribute to deviations as large as the total error residuals. Conversely, at locations P02, P07, and P08, modeled depths fluctuated by 4~cm or less throughout the entire 7-hour period, indicating that error metrics at these points are largely insensitive to timestamp ambiguity.

Overall catchment volume balance confirms that the closed computational domain retains the vast majority of incoming precipitation directly on the surface. Total precipitation contributed 7,989,142 m\textsuperscript{3} of fluid volume, while parameterized infiltration abstraction removed 816,312 m\textsuperscript{3} (accounting for 10.2\% of rainfall), as the model lacks boundary discharges, pump stations, or subterranean pipe conduits through which mass could evacuate. Final surface storage at the end of the simulation totaled 7,935,230 m\textsuperscript{3}, which exactly matches cumulative precipitation minus infiltration losses (7,172,830 m\textsuperscript{3}) added towards the initial antecedent storage of 762,291 m\textsuperscript{3}, with a maximum mass closure error of only 109 m\textsuperscript{3}. Although this mass balance confirms numerical volume conservation within the solver, it cannot validate localized street-scale inundation depths, and the overwhelming dominance of ponded surface storage confirms that modeled depths depend predominantly on total rainfall depth and terrain geometry.

\begin{figure}[!htbp]
\centering
\includegraphics[width=0.9\linewidth]{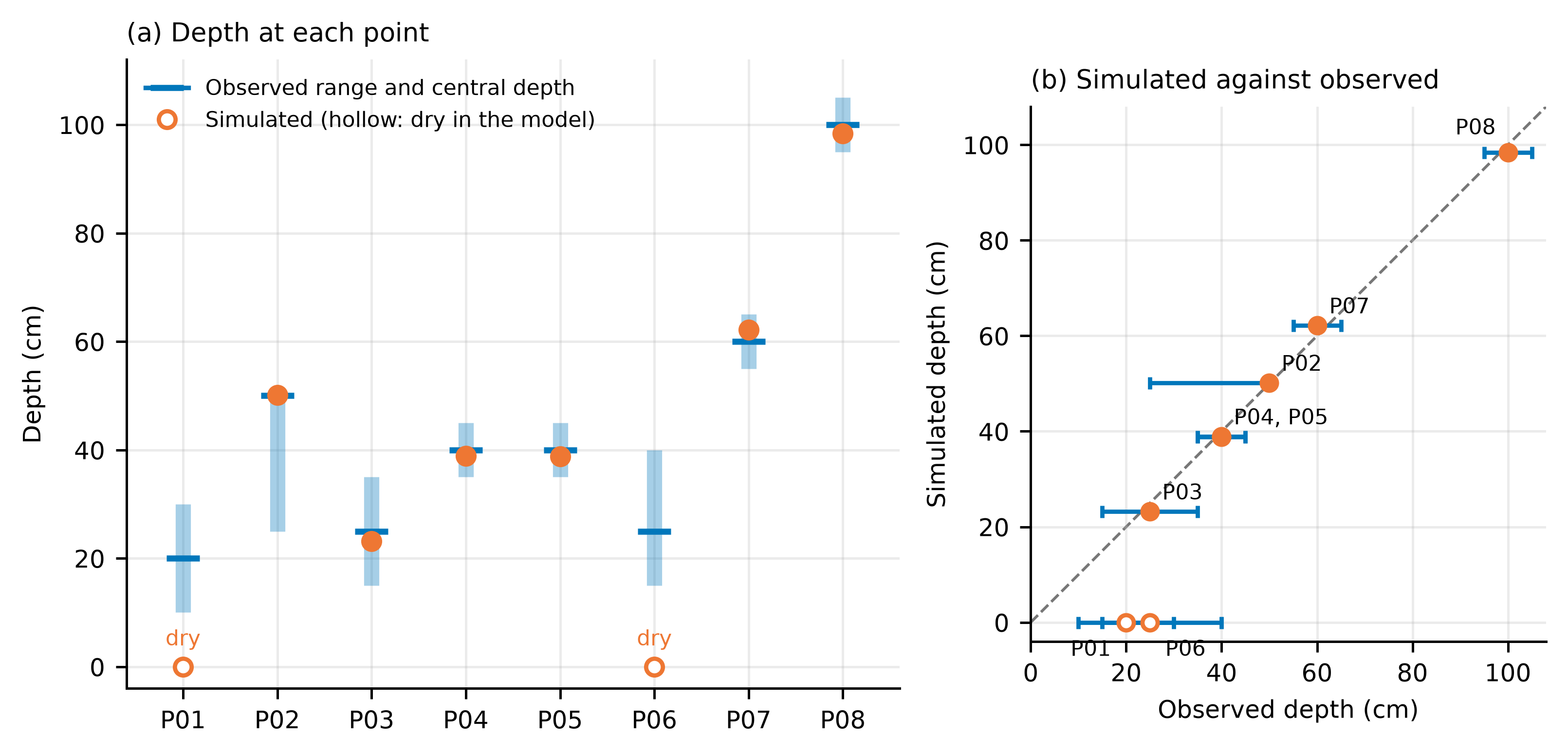}
\caption*{\textbf{Figure 6.} Observed and simulated depths at the eight calibration points. Panel (a) shows each observed range with its central value alongside the modeled depth; panel (b) plots observed against modeled depth with the 1:1 line.}
\end{figure}

\subsection{Failure case analysis and topographic artifacts}

The hydrodynamic model remained completely dry at Road 70 (P06) and Tây Mỗ Road (P01) across all 15 simulation output steps, resulting in underestimation errors of 25~cm and 20~cm that constitute the largest residuals in the final run. Notably, both failed locations are situated within the contiguous Tây Mỗ and Đại Mỗ district.

The most plausible physical explanation for the failure at Road 70 involves substantial topographic modifications occurring after the 2017 national elevation survey. The 2017 DEM contains agricultural ponds and natural retention depressions that were subsequently backfilled and developed with residential buildings prior to October 2025. In the simulation, these obsolete depression features act as artificial retention sinks that capture overland runoff that modern paved surfaces would otherwise shed into street corridors. Nevertheless, this interpretation remains a hypothesis, as no simulation utilizing updated post-development terrain was executed to demonstrate that filled ponds alone produced the dry outcome. A second compounding factor is satellite precipitation underestimation, given that the IMERG Late product recorded 47\% less rainfall than the Mễ Trì ground meteorological station over the storm window (section 3.4). Since infiltration parameters across the domain were calibrated under this depleted hyetograph, model performance cannot decouple rainfall bias from localized terrain artifacts. Future investigations incorporating both corrected contemporary terrain surfaces and bias-adjusted precipitation forcing would be required to isolate these competing failure mechanisms.

\begin{figure}[!htbp]
\centering
\includegraphics[width=0.9\linewidth]{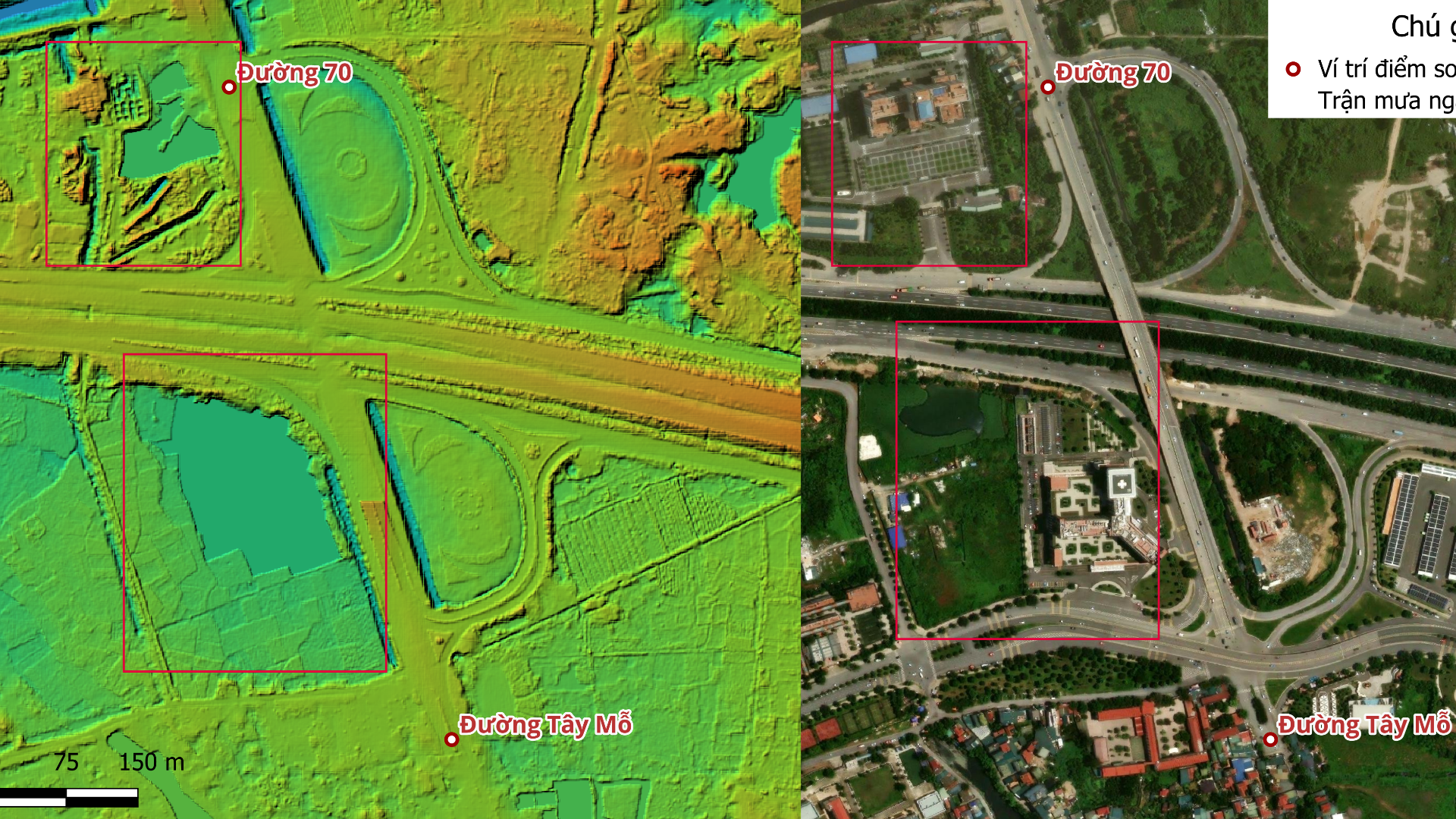}
\caption*{\textbf{Figure 7.} Terrain and imagery around the dry points P06 (Road 70) and P01 (Tây Mỗ Road). The left panels show the satellite imagery; the right panels show the conditioned DEM.}
\end{figure}

\section{Discussion}

\subsection{Physical interpretation of calibration fit and equifinality}

While simulated depths at the six inundated locations reproduce observed central values closely, this correspondence serves primarily as an empirical verification of the modeling workflow rather than evidence of predictive skill. Specifically, modeled depths align within 2.2~cm of reported central benchmarks, exhibiting a mean bias of $-$0.57~cm, an MAE of 1.34~cm, and an RMSE of 1.49~cm. However, because crowdsourced observational uncertainty spans $\pm$5 to $\pm$10~cm, an error residual of this magnitude sits entirely within observation noise and cannot be statistically separated from it. All eight observation points were utilized during parameter calibration across up to twelve adjustable values, with no independent validation points withheld. Consequently, this alignment demonstrates that a two-dimensional hydrodynamic model integrating a conditioned 2017 elevation surface, land-cover infiltration, and satellite precipitation can be calibrated towards a limited set of eyewitness observations. It does not prove that the identical parameterization would maintain predictive accuracy across unobserved metropolitan catchments or during a subsequent storm event.

In fact, the two peripheral dry locations provide substantially greater diagnostic insight than the six successfully fitted points, as they represent the sole locations where model structural deficiencies become visible. At Tây Mỗ Road (P01, 20~cm) and Road 70 (P06, 25~cm) the model simulated zero ponding, with these underestimation errors of $-$20~cm and $-$25~cm driving aggregate error metrics across the domain (bias of $-$6.05~cm, MAE of 6.63~cm, and RMSE of 11.39~cm). Three distinct physical hypotheses are consistent with this observed failure, each implying a different empirical outcome under controlled testing. To begin with, if agricultural ponds present in the 2017 elevation surface that were subsequently reclaimed act as artificial retention sinks, substituting an updated contemporary surface should generate realistic water depths at P06 while leaving the six wet locations unaffected. Secondly, if the IMERG Late hyetograph is systematically depleted (falling 47\% below the Mễ Trì meteorological station), scaling precipitation upward would wet both dry locations, but would also artificially elevate depths at the six calibrated points unless infiltration capacities are simultaneously re-adjusted. Last but not least, if unmapped stormwater conduits, culverts, or localized surface runoff concentrate overland flow along these two road corridors, a lumped rainfall forcing coupled with uniform infiltration cannot resolve localized ponding, meaning that neither rainfall adjustment nor terrain updates alone would eliminate the error.

Overall system volume balance demonstrates why physical interpretation of these calibrated loss parameters must remain exceptionally cautious. With the closed computational domain retaining approximately 90\% of incoming precipitation volume on the surface and infiltration abstracting only 10.2\%, simulated street inundation depths depend predominantly on total precipitation depth alongside the minimal abstraction capacity assigned to impervious urban surfaces. The fitted infiltration rate of 1~mm/h assigned across built-up terrain (which functions as a bulk surrogate for subsurface sewer drainage) falls substantially below the 6 to 40~mm/h effective drainage rates documented across comparable urban modeling literature. Due to the likelihood that satellite precipitation is 47\% low, this minimal drainage loss rate is consistent with an equifinality offset. In other words, a depleted precipitation input combined with minimal drainage loss produces an identical ponded surface volume as a higher rainfall total paired with realistic sewer conveyance. The calibration fit across the six wet locations is compatible with this mathematical compensation, but it is equally consistent with accurate rainfall coupled with true baseline retention, and available monitoring data cannot distinguish between these competing explanations.

\subsection{Operational data scarcity limits and future priorities}

Methodological limitations of this investigation fall into three broad categories: factors that weaken physical inference from the calibration fit, factors that obscure the physical meaning of model parameters, and constraints that limit external replicability.

Model calibration against crowdsourced media depths in the absence of an independent validation dataset represents the constraint with the most severe structural consequence. All eight depth observations originate from only two journalistic sources (six from Tuổi Trẻ and two from Tiền Phong), with depth ranges established through subjective visual heuristics rather than standardized physical surveying. Reported observation timestamps are highly uneven, and two official entries specify only a broad ``afternoon'' window. Furthermore, as every available observation point was utilized during parameter tuning and intermediate calibration trials were not systematically recorded, this calibration fit cannot differentiate a physically sound representation from an over-parameterized model. The optimization trajectory linking baseline literature parameters to final calibrated values cannot be reconstructed, meaning that readers can evaluate only the final parameter configuration.

The mathematical trade-off between satellite precipitation forcing and surface infiltration abstraction fundamentally constrains the physical interpretation of hydraulic parameters. Precipitation forcing relies entirely on a single satellite series discretized across an 11~km grid at 30-minute intervals, with no local rain gauge network available to validate IMERG Late performance during this typhoon. Available meteorological station accumulations suggest that the satellite product substantially underestimates extreme convective intensities. Calibration inevitably absorbs this negative precipitation bias directly into the land-cover infiltration rates, as both terms regulate total surface water volume. Consequently, calibrated loss rates represent effective empirical coefficients specific to this satellite hyetograph, and they should not be transferred to different storm events or alternative precipitation products. While sensitivity simulations with scaled rainfall could evaluate this interaction, such parameter sweeps were not conducted in the present study.

Additional structural boundaries involve the boundary schematization and digital elevation model. The computational domain is hydrologically closed, containing no sewer network, pump stations, or external boundary discharges, meaning that the Nhuệ River conveys zero external upstream flow. Any vertical discrepancy in this baseline stage alters available open-water storage capacity, directly influencing overland spilling onto adjacent street corridors. Furthermore, because the digital elevation model dates from 2017, all urban infrastructure constructed during the intervening eight years is omitted, leaving reclaimed agricultural ponds as artificial topographic depressions. Finally, channel bathymetry was surveyed approximately five months following the disaster, requiring the unverified assumption that bed morphology remained unaltered across the intervening period.

\section{Conclusion}

In conclusion, the present study constructed and evaluated a 32.8 km\textsuperscript{2} 3Di subgrid hydrodynamic model across Mỹ Đình and Tây Mỗ for the flood disaster following Typhoon Matmo on 7 October 2025. Under severe conditions of data scarcity, the modeling framework integrated a systematically repaired 2017 national elevation model, parameterized infiltration and roughness rasters assigned across satellite land-cover classifications, and a single lumped IMERG Late satellite precipitation hyetograph. Hydrodynamic parameters were subsequently calibrated against eight crowdsourced depth observations retrieved from news media reports. The simulation results demonstrated that six observation points held surface water and reproduced reported central values within 2.2~cm, falling directly inside the $\pm$5 to $\pm$10~cm observational uncertainty intervals. However, the model remained completely dry at Tây Mỗ Road and Road 70, resulting in underestimation errors of $-$20~cm and $-$25~cm that raised overall root mean square error across all eight points to 11.39~cm. Crucially, this correspondence represents a calibration fit rather than an unconstrained predictive validation. Because all eight observation points were utilized to adjust numerical parameters, no independent validation dataset exists, while the closed domain boundary and the mathematical trade-off between satellite precipitation and surface infiltration leave the physical meaning of calibrated loss rates open.

Altogether, our findings demonstrate the feasibility of the workflow for simulating extreme inundation in unmonitored metropolitan environments, while also highlighting structural boundaries that require further investigation. What remains to be tested is whether this calibration fit survives when subjected to alternative precipitation forcing and an independent set of surveyed ground depths. Specifically, a numerical simulation driven by bias-corrected radar precipitation, a comparative run utilizing the raw unconditioned terrain model, and a contemporary ground survey of reclaimed agricultural ponds would effectively decouple the three competing hypotheses for the dry failure points. None of these diagnostic experiments was feasible with the field data available to the present study. Without such validation data, the current model supports the proposed workflow as a means of reproducing a small set of reported flood depths in data-sparse urban catchments, but does not substantiate claims of generalized predictive skill. Future work should examine parallel simulations across neighboring Hanoi drainage basins, as canals, shared pumping stations, and backwater levels link these basins physically. Running each catchment as a separate model and exchanging boundary flows at each time step would keep computation manageable while tracking cross-basin water transfers.

\clearpage

\section*{Funding}

TECOTEC TECHNOLOGIES JOINT STOCK COMPANY provided primary institutional funding for the execution of this research.

\section*{Ethics declaration}

The present study utilizes no human-subject clinical data and no personally identifiable information. Inundation depth records were extracted exclusively from publicly accessible digital news articles and function strictly as reported water depths at designated geographical locations. Original media photographs and accompanying journalistic text remain the intellectual property of the respective news agencies that published them (Appendix A).

\begin{landscape}
\section*{Appendix A. Observations}
\begin{table}[!htbp]
\centering
\scriptsize
\setlength{\tabcolsep}{3pt}
\caption*{Table A1. Depth observations used to calibrate the model.}
\begin{tabular}{>{\raggedright\arraybackslash}p{0.030\dimexpr(\paperheight-5cm)-60pt\relax}>{\raggedright\arraybackslash}p{0.181\dimexpr(\paperheight-5cm)-60pt\relax}>{\raggedright\arraybackslash}p{0.181\dimexpr(\paperheight-5cm)-60pt\relax}>{\raggedright\arraybackslash}p{0.050\dimexpr(\paperheight-5cm)-60pt\relax}>{\raggedright\arraybackslash}p{0.181\dimexpr(\paperheight-5cm)-60pt\relax}>{\centering\arraybackslash}p{0.050\dimexpr(\paperheight-5cm)-60pt\relax}>{\centering\arraybackslash}p{0.050\dimexpr(\paperheight-5cm)-60pt\relax}>{\centering\arraybackslash}p{0.050\dimexpr(\paperheight-5cm)-60pt\relax}>{\raggedright\arraybackslash}p{0.045\dimexpr(\paperheight-5cm)-60pt\relax}>{\raggedright\arraybackslash}p{0.181\dimexpr(\paperheight-5cm)-60pt\relax}}
\toprule
ID & Place & Time (UTC+7) & Source & Reference object & Minimum (cm) & Central (cm) & Maximum (cm) & Grade & Link \\
\midrule
P01 & Tây Mỗ Road & 7 October 2025 08:20 & Tuổi Trẻ & Electric motorbike with two riders, water at the lower wheel & 10 & 20 & 30 & medium & \url{https://cdn2.tuoitre.vn/thumb_w/640/471584752817336320/2025/10/7/base64-1759801610076696415705.jpeg} \\
P02 & National Exhibition Center for Planning and Architecture & 7 October 2025 09:20 & Tuổi Trẻ & People in the forecourt, water from the shin to the knee & 25 & 50 & 50 & low & \url{https://tuoitre.vn/mua-xuyen-dem-duong-pho-ha-noi-bat-dau-ngap-sau-20251007070931343.htm#content-8} \\
P03 & Phú Đô village gate & 7 October 2025 08:20 & Tuổi Trẻ & Car tires and a bicycle wheel, water over the lower tire & 15 & 25 & 35 & medium & \url{https://cdn2.tuoitre.vn/thumb_w/640/471584752817336320/2025/10/7/base64-1759800690025228681092.jpeg} \\
P04 & Lê Đức Thọ- Tân Mỹ Street & 7 October 2025 08:17 & Tuổi Trẻ & Stated in the article & 35 & 40 & 45 & medium & \url{https://tuoitre.vn/mua-xuyen-dem-duong-pho-ha-noi-bat-dau-ngap-sau-20251007070931343.htm} \\
P05 & Lê Quang Đạo Street & 7 October 2025 08:17 & Tuổi Trẻ & Stated in the article & 35 & 40 & 45 & medium & \url{https://tuoitre.vn/mua-xuyen-dem-duong-pho-ha-noi-bat-dau-ngap-sau-20251007070931343.htm} \\
P06 & Road 70 & 7 October 2025 09:20 & Tuổi Trẻ & People wading and pulling carts with motorbikes, water at the lower leg to knee & 15 & 25 & 40 & medium & \url{https://cdn2.tuoitre.vn/thumb_w/640/471584752817336320/2025/10/7/base64-17598038673961217478297.jpeg} \\
P07 & Km8+200: Thang Long Boulevard & 7 October 2025 15:00 (article says afternoon) & Tiền Phong & Stated in the article & 55 & 60 & 65 & medium & \url{https://tienphong.vn/ha-noi-con-tren-100-diem-ngap-co-23-diem-khong-the-luu-thong-post1784918.tpo} \\
P08 & Km4+944: Thang Long Boulevard, service underpass of the left-bank Nhuệ dike & 7 October 2025 15:00 (article says afternoon) & Tiền Phong & Stated in the article & 95 & 100 & 105 & medium & \url{https://tienphong.vn/ha-noi-con-tren-100-diem-ngap-co-23-diem-khong-the-luu-thong-post1784918.tpo} \\
\bottomrule
\end{tabular}
\end{table}
\end{landscape}

\section*{Appendix B. Model details}

\subsection*{DEM repair steps}

\paragraph{}\textbf{Repair 1: the Nhuệ River bed with survey points.}

\begin{enumerate}

\item Draw the polygon of the water body, and convert its outline to a line.

\item Create points along the line at intervals of 1~m.

\item Read the DEM elevation at each point from a buffer of less than 1~m inside the polygon, so that the value is the elevation at the water's edge.

\item Merge the bank points with the 17,386 survey points. The dataset holds 921 points that repeat the x and y position of another point, with little effect on the interpolated bed.

\item Interpolate a first DEM with Kriging.

\item Draw the thalweg through the deepest part of the first DEM. Create points along it at the lowest elevation, and merge them with the earlier points.

\item Interpolate a second DEM with Kriging at 1~m resolution from the bank points, the survey points, and the thalweg points.

\item Repeat steps 6 and 7. The number of repeats was not recorded.

\item Merge the bed model into the full DEM.

\end{enumerate}

\textbf{Repair 2: lakes and ponds without survey data.} The survey covers the Nhuệ River bed only, so every lake and pond is repaired this way. For each such water body:

\begin{enumerate}

\item Digitize the boundary from the DEM and from recent satellite imagery.

\item Assign the DEM elevation at the edge of the lake to the boundary.

\item Create buffer rings that shrink toward the center.

\item Assign each ring an estimated bed elevation, chosen by judgment.

\item Interpolate with Kriging, and clip the result to the boundary.

\end{enumerate}

\textbf{Repair 3: underpasses.} Two underpasses on Thang Long Boulevard are repaired, the Trung Hòa underpass and the underpass at Km4+944 (the service underpass of the left-bank Nhuệ dike). For each underpass:

\begin{enumerate}

\item Draw 4 profiles across and along the road, and export distance and DEM values.

\item Delete the DEM values that describe the deck or the upper road.

\item Draw a control polygon around the underpass surface.

\item Interpolate the surface inside the polygon by inverse distance weighting, along the natural slope of the road.

\item Merge the patch into the full DEM.

\end{enumerate}

\textbf{Repair 4: the Mỹ Đình Pearl pit.} A total of 5,639 cells lie within 0.02~m of the absolute elevation minimum of $-$0.61~m. The repair procedure interpolates the surface utilizing the inverse distance weighting method adopted for road underpasses. No artificial target elevation was assigned manually, with the conditioned surface adopting boundary elevations from undisturbed terrain surrounding the excavation. Consequently, the entire spatial patch comprising approximately 23,000 cells was elevated by an average of 5.9~m.

\subsection*{Rainfall input generation steps}

\begin{enumerate}

\item Download the 131 granules of IMERG Late, version 07B, for the bounding box of the model area and the event window.

\item Keep each 0.1-degree cell that overlaps the bounding box. Four cells overlap it, and their weights are 45.4\%, 53.3\%, 0.6\%, and 0.7\%.

\item Compute the weight of each cell as the fraction of the cell inside the bounding box in longitude, multiplied by the fraction in latitude. Skip cells that hold the fill value (-9999.9). Take the weighted mean of the rate over the cells.

\item Multiply each half-hourly rate in mm/h by 0.5~h to get the depth in mm for the 30-minute step.

\item Write two columns, the minutes since the start of the event and the depth for the step. 3Di reads the second column as a depth per step, not as a rate.

\end{enumerate}

The series starts at minute 0 and ends at minute 3,900.

\subsection*{Friction and infiltration rasters}

The background Manning's roughness coefficient of 0.013 covers 54.3\% of the computational domain, encompassing paved road networks alongside 43.8\% of built-up residential cells. The value of 0.036 covers 29.2\% of the active cells. Prior to model calibration, the authors assigned a baseline roughness of 0.035 to bare soil and active construction sites. Because bare soil occupies only 3.8\% of the ESA WorldCover classification, additional transitional land features were assigned this intermediate friction parameter. Furthermore, the elevated roughness value of 0.060 covers 13.4\% of the cells, representing urban residential clusters alongside agricultural paddy land. The open water friction value of 0.020 covers 2.7\% of the domain, corresponding primarily to retention lakes and community ponds. Finally, a Manning's n of 0.025 covers 0.5\% of the domain, delineating the incised bed of the Nhuệ River and its primary tributary channels. These spatial roughness rasters were constructed by combining independent geospatial layers for land cover, roadway alignments, building footprints, and surface water bodies. For each categorical class within the source datasets, corresponding values for Manning's n and infiltration capacity were assigned via raster calculation tables in QGIS. Because the raw ESA WorldCover layer is resolved at 10~m while hydrodynamic rasters are discretized at 1~m, land-cover classes were resampled to the 1~m model grid using nearest-neighbor interpolation. Nearest-neighbor resampling preserves categorical class integers without smoothing, meaning that cell-by-cell parameter attribution remains mathematically identical regardless of processing sequence.

\begin{figure}[!htbp]
\centering
\includegraphics[width=0.9\linewidth]{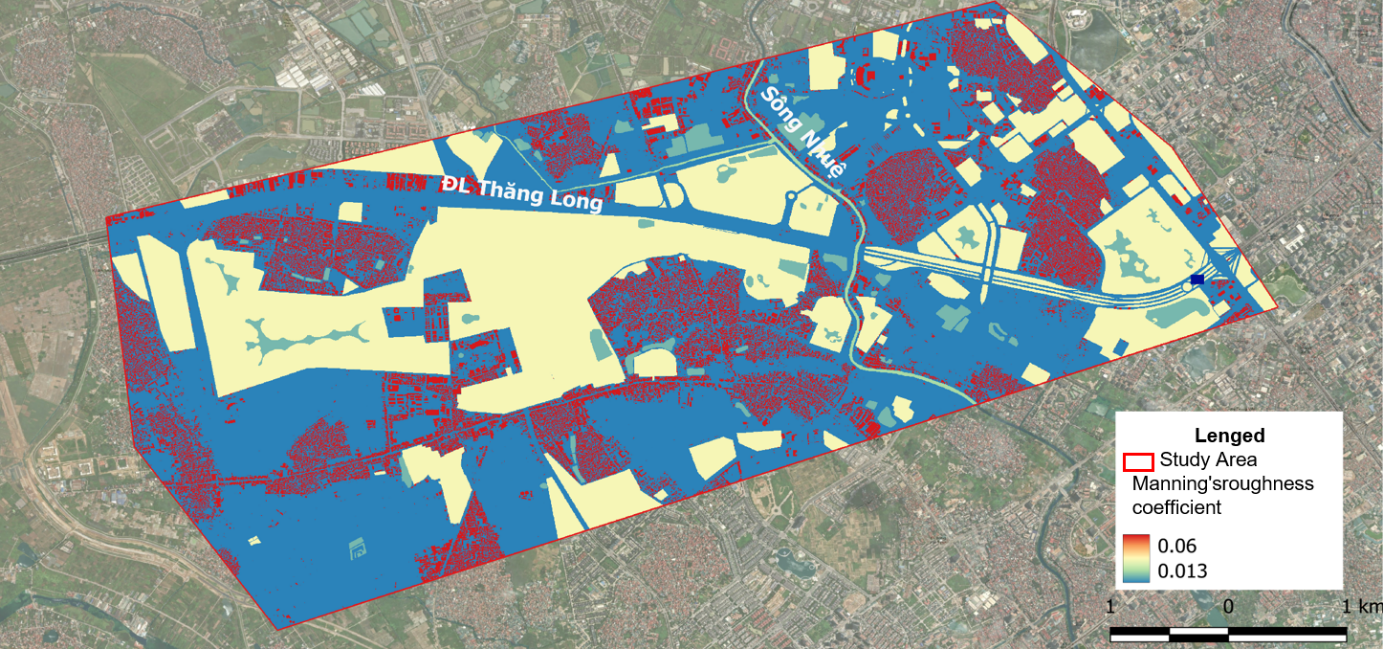}
\caption*{\textbf{Figure B1.} Manning's roughness coefficient (n) raster used in the model.}
\end{figure}

\begin{figure}[!htbp]
\centering
\includegraphics[width=0.9\linewidth]{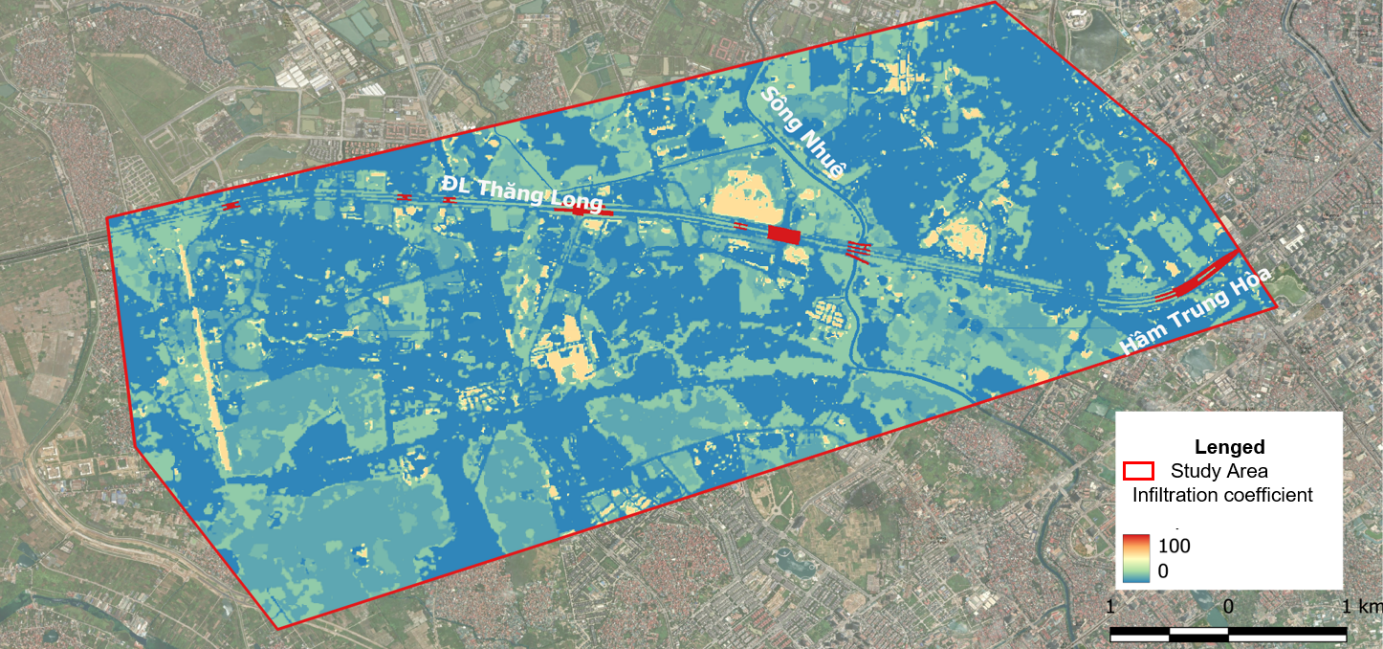}
\caption*{\textbf{Figure B2.} Infiltration raster used in the model (mm/h).}
\end{figure}

\begin{figure}[!htbp]
\centering
\includegraphics[width=0.9\linewidth]{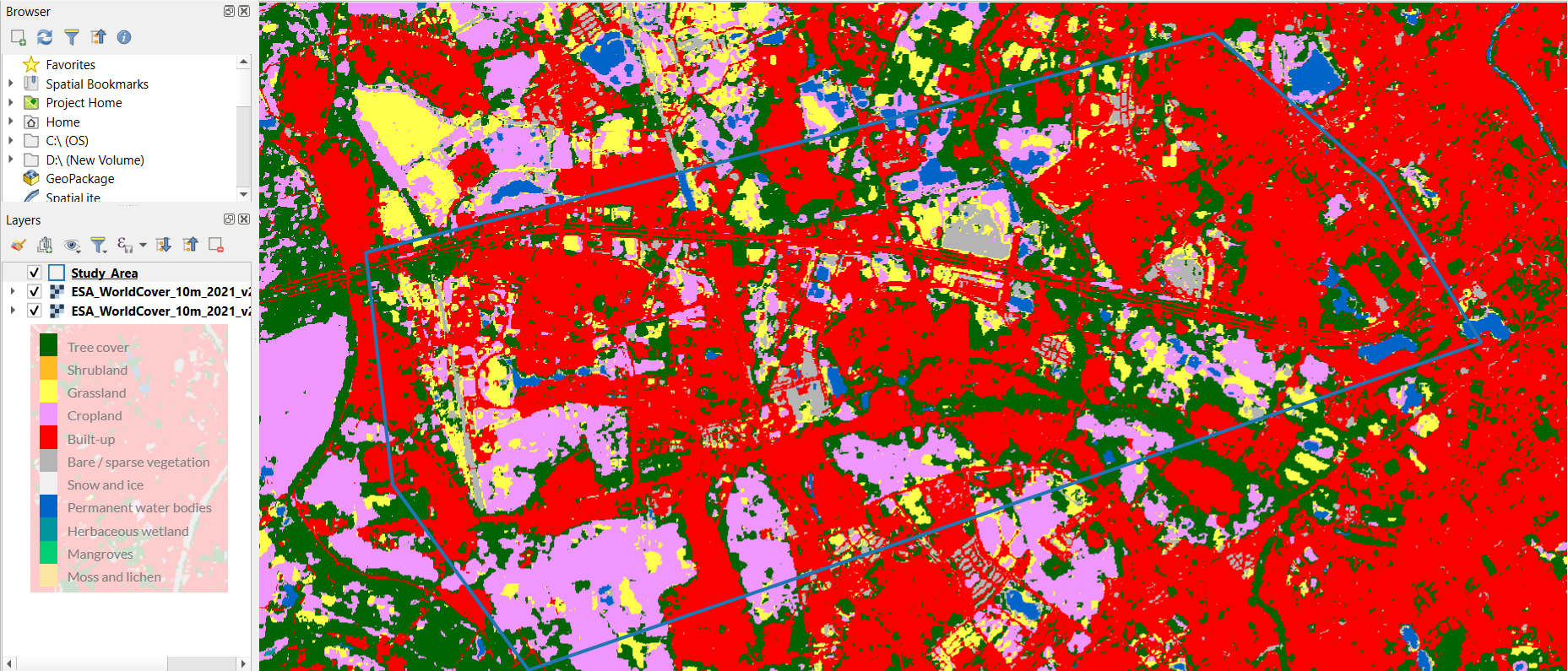}
\caption*{\textbf{Figure B3.} ESA WorldCover land cover (10~m, 2021) used to build the infiltration raster.}
\end{figure}

\FloatBarrier

\section*{References}

\begingroup\small\setlength{\parindent}{0pt}\setlength{\parskip}{0.5em}\raggedright

\hangindent=1.5em\hangafter=1 Abdullah, A. F., Vojinovic, Z., Price, R. K., \& Aziz, N. A. A. (2012). Improved methodology for processing raw LiDAR data to support urban flood modelling -- accounting for elevated roads and bridges. \emph{Journal of Hydroinformatics}, \emph{14}(2), 253-269. https://doi.org/10.2166/hydro.2011.009

\hangindent=1.5em\hangafter=1 Alizadeh Kharazi, B., \& Behzadan, A. H. (2021). Flood depth mapping in street photos with image processing and deep neural networks. \emph{Computers, Environment and Urban Systems}, \emph{88}, 101628. https://doi.org/10.1016/j.compenvurbsys.2021.101628

\hangindent=1.5em\hangafter=1 Amini, E., Marsooli, R., Orton, P., Kasaei, S., \& Hajj, M. (2026). High resolution simulation of urban compound flooding under climate impacts. \emph{Urban Climate}, \emph{65}, 102771. https://doi.org/10.1016/j.uclim.2026.102771

\hangindent=1.5em\hangafter=1 Apel, H., Benisch, J., Helm, B., Vorogushyn, S., \& Merz, B. (2024). Fast urban inundation simulation with RIM2D for flood risk assessment and forecasting. \emph{Frontiers in Water}, \emph{6}, 1310182. https://doi.org/10.3389/frwa.2024.1310182

\hangindent=1.5em\hangafter=1 Casulli, V. (2009). A high-resolution wetting and drying algorithm for free-surface hydrodynamics. \emph{International Journal for Numerical Methods in Fluids}, \emph{60}(4), 391-408. https://doi.org/10.1002/fld.1896

\hangindent=1.5em\hangafter=1 Casulli, V., \& Stelling, G. S. (2011). Semi-implicit subgrid modelling of three-dimensional free-surface flows. \emph{International Journal for Numerical Methods in Fluids}, \emph{67}(4), 441-449. https://doi.org/10.1002/fld.2361

\hangindent=1.5em\hangafter=1 Chai, T., \& Draxler, R. R. (2014). Root mean square error (RMSE) or mean absolute error (MAE)? -- Arguments against avoiding RMSE in the literature. \emph{Geoscientific Model Development}, \emph{7}(3), 1247-1250. https://doi.org/10.5194/gmd-7-1247-2014

\hangindent=1.5em\hangafter=1 Chinhphu.vn (2025). \emph{Tình hình MƯA, NGẬP ÚNG tại HÀ NỘI; cảnh báo MƯA LỚN, nguy cơ LŨ QUÉT, SẠT LỞ ĐẤT ở BẮC BỘ, THANH HÓA}. Government news portal, published 7 October 2025 12:06 (UTC+7). Reports a bulletin of the national center for hydro-meteorological forecasting. https://xaydungchinhsach.chinhphu.vn/cap-nhat-tinh-hinh-mua-ngap-ung-tai-ha-noi-canh-bao-mua-lon-o-bac-bo-thanh-hoa-119251007083105629.htm

\hangindent=1.5em\hangafter=1 Chow, V. T. (1959). \emph{Open-channel hydraulics}. McGraw-Hill.

\hangindent=1.5em\hangafter=1 Environment Agency (2019). \emph{What is the risk of flooding from surface water map? Version 2.0}. Grey literature, no DOI. Not read. Cited through Ming et al. (2025), Table 1. Environment Agency.

\hangindent=1.5em\hangafter=1 Fohringer, J., Dransch, D., Kreibich, H., \& Schröter, K. (2015). Social media as an information source for rapid flood inundation mapping. \emph{Natural Hazards and Earth System Sciences}, \emph{15}(12), 2725-2738. https://doi.org/10.5194/nhess-15-2725-2015

\hangindent=1.5em\hangafter=1 Gourou, P. (1936). \emph{Les Paysans du Delta Tonkinois: Étude de géographie humaine}. Éditions d'Art et d'Histoire. (English translation: \emph{The Peasants of the Tonkin Delta}, Human Relations Area Files, 1955).

\hangindent=1.5em\hangafter=1 Huffman, G., Stocker, E., Bolvin, D., Nelkin, E., \& Tan, J. (2023). \emph{GPM IMERG Late Precipitation L3 Half Hourly 0.1 degree x 0.1 degree V07}. NASA Goddard Earth Sciences Data and Information Services Center (GES DISC). https://doi.org/10.5067/GPM/IMERG/3B-HH-L/07

\hangindent=1.5em\hangafter=1 Li, D., Hou, J., Shen, R., Li, B., Tong, Y., \& Wang, T. (2023). Approximation method for the sewer drainage effect for urban flood modeling in areas without drainage-pipe data. \emph{Frontiers in Environmental Science}, \emph{11}, 1134985. https://doi.org/10.3389/fenvs.2023.1134985

\hangindent=1.5em\hangafter=1 Li, D., Hou, J., Xia, J., Tong, Y., Yang, D., Zhang, D., \& Gao, X. (2020). An Efficient Method for Approximately Simulating Drainage Capability for Urban Flood. \emph{Frontiers in Earth Science}, \emph{8}, 159. https://doi.org/10.3389/feart.2020.00159

\hangindent=1.5em\hangafter=1 Loftis, J. D., Wang, H. V., Hamilton, S. E., \& Forrest, D. R. (2014). \emph{Combination of Lidar Elevations, Bathymetric Data, and Urban Infrastructure in a Sub-Grid Model for Predicting Inundation in New York City during Hurricane Sandy}. arXiv preprint. https://doi.org/10.48550/arXiv.1412.0966

\hangindent=1.5em\hangafter=1 Luo, P., Mu, D., Xue, H., Ngo-Duc, T., Dang-Dinh, K., Takara, K., Nover, D., \& Schladow, G. (2018). Flood inundation assessment for the Hanoi Central Area, Vietnam under historical and extreme rainfall conditions. \emph{Scientific Reports}, \emph{8}(1), 12623. https://doi.org/10.1038/s41598-018-30024-5

\hangindent=1.5em\hangafter=1 Lê, B. T. (2009). \emph{Thiên nhiên Việt Nam} [Nature of Vietnam] (6th ed.). NXB Giáo dục Việt Nam.

\hangindent=1.5em\hangafter=1 Ming, X., Liang, Q., \& Jiang, J. (2025). Large-scale high-resolution hydrodynamic modelling of urban floods: Some practical considerations. \emph{Journal of Hydro-environment Research}, \emph{59}, 100655. https://doi.org/10.1016/j.jher.2025.100655

\hangindent=1.5em\hangafter=1 Neal, J., Hawker, L., Savage, J., Durand, M., Bates, P., \& Sampson, C. (2021). Estimating River Channel Bathymetry in Large Scale Flood Inundation Models. \emph{Water Resources Research}, \emph{57}(5), e2020WR028301. https://doi.org/10.1029/2020WR028301

\hangindent=1.5em\hangafter=1 Nguyen, V. D., Rouzegari, N., Dao, V., Almutlaq, F., Nguyen, P., \& Sorooshian, S. (2025). Comparative Analysis of Satellite-Based Precipitation Products During Extreme Rainfall from Super Typhoon Yagi in Hanoi, Vietnam (September 2024). \emph{Remote Sensing}, \emph{17}(9), 1598. https://doi.org/10.3390/rs17091598

\hangindent=1.5em\hangafter=1 Platzek, F. W., Stelling, G. S., Jankowski, J. A., Patzwahl, R., \& Pietrzak, J. D. (2015). An efficient semi-implicit subgrid method for free-surface flows on hierarchical grids. \emph{International Journal for Numerical Methods in Fluids}, \emph{80}(12), 715-741. https://doi.org/10.1002/fld.4172

\hangindent=1.5em\hangafter=1 Rosser, J. F., Leibovici, D. G., \& Jackson, M. J. (2017). Rapid flood inundation mapping using social media, remote sensing and topographic data. \emph{Natural Hazards}, \emph{87}(1), 103-120. https://doi.org/10.1007/s11069-017-2755-0

\hangindent=1.5em\hangafter=1 Roversi, G., Pancaldi, M., Cossich, W., Corradini, D., Nguyen, T. T. N., Nguyen, T. V., \& Porcù, F. (2024). The Extreme Rainfall Events of the 2020 Typhoon Season in Vietnam as Seen by Seven Different Precipitation Products. \emph{Remote Sensing}, \emph{16}(5), 805. https://doi.org/10.3390/rs16050805

\hangindent=1.5em\hangafter=1 Stelling, G. S. (2012). Quadtree flood simulations with sub-grid digital elevation models. \emph{Proceedings of the Institution of Civil Engineers - Water Management}, \emph{165}(10), 567-580. https://doi.org/10.1680/wama.12.00018

\hangindent=1.5em\hangafter=1 Su, Y. F., Lin, Y. T., Jang, J. H., \& Han, J. Y. (2022). High-Resolution Flood Simulation in Urban Areas Through the Application of Remote Sensing and Crowdsourcing Technologies. \emph{Frontiers in Earth Science}, \emph{9}, 756198. https://doi.org/10.3389/feart.2021.756198

\hangindent=1.5em\hangafter=1 Thanh Niên (2025). \emph{Hà Nội mưa lớn sau bão Matmo, nhiều tuyến phố đã ngập sâu}. News article, published 7 October 2025 06:55 (UTC+7). https://thanhnien.vn/ha-noi-mua-lon-sau-bao-matmo-nhieu-tuyen-pho-da-ngap-sau-185251007063653491.htm

\hangindent=1.5em\hangafter=1 Thủ tướng Chính phủ (2013). \emph{Quyết định số 725/QĐ-TTg ngày 10/5/2013 phê duyệt Quy hoạch thoát nước Thủ đô Hà Nội đến năm 2030 và tầm nhìn đến năm 2050}. Decision of the Prime Minister of Vietnam. Not verifiable by DOI. https://thuvienphapluat.vn/van-ban/Xay-dung-Do-thi/Quyet-dinh-725-QD-TTg-nam-2013-phe-duyet-Quy-hoach-thoat-nuoc-Thu-do-Ha-Noi-186602.aspx

\hangindent=1.5em\hangafter=1 Tiền Phong (2025). \emph{Hà Nội còn trên 100 điểm ngập, có 23 điểm không thể lưu thông}. News article by Anh Trọng, published 7 October 2025 19:28 (UTC+7). https://tienphong.vn/ha-noi-con-tren-100-diem-ngap-co-23-diem-khong-the-luu-thong-post1784918.tpo

\hangindent=1.5em\hangafter=1 Tuổi Trẻ (2025). \emph{Mưa xuyên đêm, nhiều tuyến phố Hà Nội thành sông}. News article, published 7 October 2025 07:10 (UTC+7). https://tuoitre.vn/mua-xuyen-dem-duong-pho-ha-noi-bat-dau-ngap-sau-20251007070931343.htm

\hangindent=1.5em\hangafter=1 van Dam, A., Kleptsova, O., Stelling, G., \& Pleumeekers, O. (2012). \emph{3Di subgrid flooding}. CRC Press. https://doi.org/10.1201/b13715-72

\hangindent=1.5em\hangafter=1 Vernimmen, R., Hooijer, A., Yuherdha, A. T., Visser, M., Pronk, M., Eilander, D., Akmalia, R., Fitranatanegara, N., Mulyadi, D., Andreas, H., Ouellette, J., \& Hadley, W. (2019). Creating a Lowland and Peatland Landscape Digital Terrain Model (DTM) from Interpolated Partial Coverage LiDAR Data for Central Kalimantan and East Sumatra, Indonesia. \emph{Remote Sensing}, \emph{11}(10), 1152. https://doi.org/10.3390/rs11101152

\hangindent=1.5em\hangafter=1 Viet Nam News (2025). Typhoon Bualoi leaves provinces in ruins. News article, published 29 September 2025. https://vietnamnews.vn/society/1726182/storm-no-10-leaves-provinces-in-ruins.html

\hangindent=1.5em\hangafter=1 VietNamNet (2025). Ba cơn bão liên tiếp tàn phá, thiên tai ngày càng khốc liệt và `dị thường'. News article. https://vietnamnet.vn/ba-con-bao-lien-tiep-tan-pha-thien-tai-ngay-cang-khoc-liet-va-di-thuong-2451670.html

\hangindent=1.5em\hangafter=1 VOV (2025). \emph{Bão số 11 Matmo suy yếu thành áp thấp nhiệt đới, miền Bắc mưa lớn diện rộng}. News article, published 6 October 2025. https://vov.gov.vn/bao-so-11-matmo-suy-yeu-thanh-ap-thap-nhiet-doi-mien-bac-mua-lon-dien-rong-dtnew-1117653

\hangindent=1.5em\hangafter=1 Wang, Y., Chen, A. S., Fu, G., Djordjević, S., Zhang, C., \& Savić, D. A. (2018). An integrated framework for high-resolution urban flood modelling considering multiple information sources and urban features. \emph{Environmental Modelling and Software}, \emph{107}, 85-95. https://doi.org/10.1016/j.envsoft.2018.06.010

\hangindent=1.5em\hangafter=1 Xu, J., Qi, Y., Li, D., \& Zhao, Z. (2024). Can IMERG QPE product capture the heavy rain on urban flood scale?. \emph{Science of The Total Environment}, \emph{933}, 173022. https://doi.org/10.1016/j.scitotenv.2024.173022

\hangindent=1.5em\hangafter=1 Yu, D., Yin, J., \& Liu, M. (2016). Validating city-scale surface water flood modelling using crowd-sourced data. \emph{Environmental Research Letters}, \emph{11}(12), 124011. https://doi.org/10.1088/1748-9326/11/12/124011

\hangindent=1.5em\hangafter=1 Zanaga, D., Van De Kerchove, R., Daems, D., De Keersmaecker, W., Brockmann, C., Kirches, G., Wevers, J., Cartus, O., Santoro, M., Fritz, S., Lesiv, M., Herold, M., Tsendbazar, N. E., Xu, P., Ramoino, F., \& Arino, O. (2022). \emph{ESA WorldCover 10~m 2021 v200}. Zenodo. v200. https://doi.org/10.5281/zenodo.7254221

\endgroup


\begin{thebibliography}{39}
\providecommand{\natexlab}[1]{#1}
\providecommand{\url}[1]{\texttt{#1}}
\expandafter\ifx\csname urlstyle\endcsname\relax
  \providecommand{\doi}[1]{doi: #1}\else
  \providecommand{\doi}{doi: \begingroup \urlstyle{rm}\Url}\fi

\bibitem[Abdullah(2012)]{abdullah2012lidar}
Abdullah (2012).
\newblock Improved methodology for processing raw LiDAR data to support urban flood modelling -- accounting for elevated roads and bridges. \emph{Journal of Hydroinformatics}, \emph{14}(2), 253-269. \url{https://doi.org/10.2166/hydro.2011.009}

\bibitem[Alizadeh Kharazi(2021)]{kharazi2021depth}
Alizadeh Kharazi (2021).
\newblock Flood depth mapping in street photos with image processing and deep neural networks. \emph{Computers, Environment and Urban Systems}, \emph{88}, 101628. \url{https://doi.org/10.1016/j.compenvurbsys.2021.101628}

\bibitem[Amini(2026)]{amini2026hoboken}
Amini (2026).
\newblock High resolution simulation of urban compound flooding under climate impacts. \emph{Urban Climate}, \emph{65}, 102771. \url{https://doi.org/10.1016/j.uclim.2026.102771}

\bibitem[Apel(2024)]{apel2024rim2d}
Apel (2024).
\newblock Fast urban inundation simulation with RIM2D for flood risk assessment and forecasting. \emph{Frontiers in Water}, \emph{6}, 1310182. \url{https://doi.org/10.3389/frwa.2024.1310182}

\bibitem[Casulli(2009)]{casulli2009wetting}
Casulli (2009).
\newblock A high-resolution wetting and drying algorithm for free-surface hydrodynamics. \emph{International Journal for Numerical Methods in Fluids}, \emph{60}(4), 391-408. \url{https://doi.org/10.1002/fld.1896}

\bibitem[Casulli(2011)]{casulli2011subgrid}
Casulli (2011).
\newblock Semi-implicit subgrid modelling of three-dimensional free-surface flows. \emph{International Journal for Numerical Methods in Fluids}, \emph{67}(4), 441-449. \url{https://doi.org/10.1002/fld.2361}

\bibitem[Chai(2014)]{chai2014rmse}
Chai (2014).
\newblock Root mean square error (RMSE) or mean absolute error (MAE)? -- Arguments against avoiding RMSE in the literature. \emph{Geoscientific Model Development}, \emph{7}(3), 1247-1250. \url{https://doi.org/10.5194/gmd-7-1247-2014}

\bibitem[Chinhphu.vn(2025)]{chinhphu2025mua}
Chinhphu.vn (2025).
\newblock \emph{Tình hình MƯA, NGẬP ÚNG tại HÀ NỘI; cảnh báo MƯA LỚN, nguy cơ LŨ QUÉT, SẠT LỞ ĐẤT ở BẮC BỘ, THANH HÓA}. Government news portal, published 7 October 2025 12:06 (UTC+7). Reports a bulletin of the national center for hydro-meteorological forecasting. \url{https://xaydungchinhsach.chinhphu.vn/cap-nhat-tinh-hinh-mua-ngap-ung-tai-ha-noi-canh-bao-mua-lon-o-bac-bo-thanh-hoa-119251007083105629.htm}

\bibitem[Chow(1959)]{chow1959open}
Chow (1959).
\newblock \emph{Open-channel hydraulics}. McGraw-Hill.

\bibitem[Environment Agency(2019)]{envagency2019}
Environment Agency (2019).
\newblock \emph{What is the risk of flooding from surface water map? Version 2.0}. Grey literature, no DOI. Not read. Cited through Ming et al. (2025), Table 1. Environment Agency.

\bibitem[Fohringer(2015)]{fohringer2015social}
Fohringer (2015).
\newblock Social media as an information source for rapid flood inundation mapping. \emph{Natural Hazards and Earth System Sciences}, \emph{15}(12), 2725-2738. \url{https://doi.org/10.5194/nhess-15-2725-2015}

\bibitem[Gourou(1936)]{gourou1936tonkin}
Gourou (1936).
\newblock \emph{Les Paysans du Delta Tonkinois: Étude de géographie humaine}. Éditions d'Art et d'Histoire. (English translation: \emph{The Peasants of the Tonkin Delta}, Human Relations Area Files, 1955).

\bibitem[Huffman(2023)]{huffman2023imerg}
Huffman (2023).
\newblock \emph{GPM IMERG Late Precipitation L3 Half Hourly 0.1 degree x 0.1 degree V07}. NASA Goddard Earth Sciences Data and Information Services Center (GES DISC). \url{https://doi.org/10.5067/GPM/IMERG/3B-HH-L/07}

\bibitem[Lê(2009)]{le2009vietnam}
Lê (2009).
\newblock \emph{Thiên nhiên Việt Nam} [Nature of Vietnam] (6th ed.). NXB Giáo dục Việt Nam.

\bibitem[Li(2020)]{li2020drainage}
Li (2020).
\newblock An Efficient Method for Approximately Simulating Drainage Capability for Urban Flood. \emph{Frontiers in Earth Science}, \emph{8}, 159. \url{https://doi.org/10.3389/feart.2020.00159}

\bibitem[Li(2023)]{li2023sewer}
Li (2023).
\newblock Approximation method for the sewer drainage effect for urban flood modeling in areas without drainage-pipe data. \emph{Frontiers in Environmental Science}, \emph{11}, 1134985. \url{https://doi.org/10.3389/fenvs.2023.1134985}

\bibitem[Loftis(2014)]{loftis2014sandy}
Loftis (2014).
\newblock \emph{Combination of Lidar Elevations, Bathymetric Data, and Urban Infrastructure in a Sub-Grid Model for Predicting Inundation in New York City during Hurricane Sandy}. arXiv preprint. \url{https://doi.org/10.48550/arXiv.1412.0966}

\bibitem[Luo(2018)]{luo2018hanoi}
Luo (2018).
\newblock Flood inundation assessment for the Hanoi Central Area, Vietnam under historical and extreme rainfall conditions. \emph{Scientific Reports}, \emph{8}(1), 12623. \url{https://doi.org/10.1038/s41598-018-30024-5}

\bibitem[Ming(2025)]{ming2025practical}
Ming (2025).
\newblock Large-scale high-resolution hydrodynamic modelling of urban floods: Some practical considerations. \emph{Journal of Hydro-environment Research}, \emph{59}, 100655. \url{https://doi.org/10.1016/j.jher.2025.100655}

\bibitem[Neal(2021)]{neal2021bathymetry}
Neal (2021).
\newblock Estimating River Channel Bathymetry in Large Scale Flood Inundation Models. \emph{Water Resources Research}, \emph{57}(5), e2020WR028301. \url{https://doi.org/10.1029/2020WR028301}

\bibitem[Nguyen(2025)]{nguyen2025yagi}
Nguyen (2025).
\newblock Comparative Analysis of Satellite-Based Precipitation Products During Extreme Rainfall from Super Typhoon Yagi in Hanoi, Vietnam (September 2024). \emph{Remote Sensing}, \emph{17}(9), 1598. \url{https://doi.org/10.3390/rs17091598}

\bibitem[Platzek(2015)]{platzek2015hierarchical}
Platzek (2015).
\newblock An efficient semi-implicit subgrid method for free-surface flows on hierarchical grids. \emph{International Journal for Numerical Methods in Fluids}, \emph{80}(12), 715-741. \url{https://doi.org/10.1002/fld.4172}

\bibitem[Rosser(2017)]{rosser2017rapid}
Rosser (2017).
\newblock Rapid flood inundation mapping using social media, remote sensing and topographic data. \emph{Natural Hazards}, \emph{87}(1), 103-120. \url{https://doi.org/10.1007/s11069-017-2755-0}

\bibitem[Roversi(2024)]{roversi2024vietnam}
Roversi (2024).
\newblock The Extreme Rainfall Events of the 2020 Typhoon Season in Vietnam as Seen by Seven Different Precipitation Products. \emph{Remote Sensing}, \emph{16}(5), 805. \url{https://doi.org/10.3390/rs16050805}

\bibitem[Stelling(2012)]{stelling2012quadtree}
Stelling (2012).
\newblock Quadtree flood simulations with sub-grid digital elevation models. \emph{Proceedings of the Institution of Civil Engineers \- Water Management}, \emph{165}(10), 567-580. \url{https://doi.org/10.1680/wama.12.00018}

\bibitem[Su(2022)]{su2022uav}
Su (2022).
\newblock High-Resolution Flood Simulation in Urban Areas Through the Application of Remote Sensing and Crowdsourcing Technologies. \emph{Frontiers in Earth Science}, \emph{9}, 756198. \url{https://doi.org/10.3389/feart.2021.756198}

\bibitem[Thanh Niên(2025)]{thanhnien2025matmo}
Thanh Niên (2025).
\newblock \emph{Hà Nội mưa lớn sau bão Matmo, nhiều tuyến phố đã ngập sâu}. News article, published 7 October 2025 06:55 (UTC+7). \url{https://thanhnien.vn/ha-noi-mua-lon-sau-bao-matmo-nhieu-tuyen-pho-da-ngap-sau-185251007063653491.htm}

\bibitem[Thủ tướng Chính phủ(2013)]{decision725}
Thủ tướng Chính phủ (2013).
\newblock \emph{Quyết định số 725/QĐ-TTg ngày 10/5/2013 phê duyệt Quy hoạch thoát nước Thủ đô Hà Nội đến năm 2030 và tầm nhìn đến năm 2050}. Decision of the Prime Minister of Vietnam. Not verifiable by DOI. \url{https://thuvienphapluat.vn/van-ban/Xay-dung-Do-thi/Quyet-dinh-725-QD-TTg-nam-2013-phe-duyet-Quy-hoach-thoat-nuoc-Thu-do-Ha-Noi-186602.aspx}

\bibitem[Tiền Phong(2025)]{tienphong2025ngap}
Tiền Phong (2025).
\newblock \emph{Hà Nội còn trên 100 điểm ngập, có 23 điểm không thể lưu thông}. News article by Anh Trọng, published 7 October 2025 19:28 (UTC+7). \url{https://tienphong.vn/ha-noi-con-tren-100-diem-ngap-co-23-diem-khong-the-luu-thong-post1784918.tpo}

\bibitem[Tuổi Trẻ(2025)]{tuoitre2025ngap}
Tuổi Trẻ (2025).
\newblock \emph{Mưa xuyên đêm, nhiều tuyến phố Hà Nội thành sông}. News article, published 7 October 2025 07:10 (UTC+7). \url{https://tuoitre.vn/mua-xuyen-dem-duong-pho-ha-noi-bat-dau-ngap-sau-20251007070931343.htm}

\bibitem[van Dam(2012)]{vandam2012threedi}
van Dam (2012).
\newblock \emph{3Di subgrid flooding}. CRC Press. \url{https://doi.org/10.1201/b13715-72}

\bibitem[Vernimmen(2019)]{vernimmen2019lidar}
Vernimmen (2019).
\newblock Creating a Lowland and Peatland Landscape Digital Terrain Model (DTM) from Interpolated Partial Coverage LiDAR Data for Central Kalimantan and East Sumatra, Indonesia. \emph{Remote Sensing}, \emph{11}(10), 1152. \url{https://doi.org/10.3390/rs11101152}

\bibitem[Viet Nam News(2025)]{vietnamnews2025bualoi}
Viet Nam News (2025).
\newblock Typhoon Bualoi leaves provinces in ruins. News article, published 29 September 2025. \url{https://vietnamnews.vn/society/1726182/storm-no-10-leaves-provinces-in-ruins.html}

\bibitem[VietNamNet(2025)]{vietnamnet2025storms}
VietNamNet (2025).
\newblock Ba cơn bão liên tiếp tàn phá, thiên tai ngày càng khốc liệt và ‘dị thường’. News article. \url{https://vietnamnet.vn/ba-con-bao-lien-tiep-tan-pha-thien-tai-ngay-cang-khoc-liet-va-di-thuong-2451670.html}

\bibitem[VOV(2025)]{vov2025matmo}
VOV (2025).
\newblock \emph{Bão số 11 Matmo suy yếu thành áp thấp nhiệt đới, miền Bắc mưa lớn diện rộng}. News article, published 6 October 2025. \url{https://vov.gov.vn/bao-so-11-matmo-suy-yeu-thanh-ap-thap-nhiet-doi-mien-bac-mua-lon-dien-rong-dtnew-1117653}

\bibitem[Wang(2018)]{wang2018integrated}
Wang (2018).
\newblock An integrated framework for high-resolution urban flood modelling considering multiple information sources and urban features. \emph{Environmental Modelling and Software}, \emph{107}, 85-95. \url{https://doi.org/10.1016/j.envsoft.2018.06.010}

\bibitem[Xu(2024)]{xu2024imerg}
Xu (2024).
\newblock Can IMERG QPE product capture the heavy rain on urban flood scale?. \emph{Science of The Total Environment}, \emph{933}, 173022. \url{https://doi.org/10.1016/j.scitotenv.2024.173022}

\bibitem[Yu(2016)]{yu2016validating}
Yu (2016).
\newblock Validating city-scale surface water flood modelling using crowd-sourced data. \emph{Environmental Research Letters}, \emph{11}(12), 124011. \url{https://doi.org/10.1088/1748-9326/11/12/124011}

\bibitem[Zanaga(2022)]{zanaga2022worldcover}
Zanaga (2022).
\newblock \emph{ESA WorldCover 10~m 2021 v200}. Zenodo. v200. \url{https://doi.org/10.5281/zenodo.7254221}

\end{thebibliography}
\end{document}